\documentclass{SCGE}
\usepackage{rotating}
\begin{document}

\begin{picture}(0,0){\rm
\put(0,-39){\makebox[160truemm][l]{\bf {\sanhao\raisebox{2pt}{.}}
Article  {\sanhao\raisebox{1.5pt}{.}}}}}
\put(0,-52){\jiuwuhao {\textcolor[rgb]{0.5,0.5,0.5}{\sf 
}}}
\end{picture}

\def\bm{\boldsymbol}
\def\dl{\displaystyle}
\def\du{\end{document}}
\def\pi{{\uppi}}
\def\Ms{M$_{\odot}$}
\def\micron{$\mu$m}
\def\arcsec{$^{\prime\prime}$}
\def\squaredeg{${deg}^{2}$}
\def\arcminute{$^{\prime}$}

\BeginPage{1} 
\EndPage{11} 
\AuthorMark{{\rm G. Y. Zhang, et al.}}  
\AuthorMarkCite{{\rm G. Y. Zhang, D. Li, et al.}} 

\title{350 \micron\ map of the Ophiuchus molecular cloud: core mass function}

\author[1]{Guoyin Zhang}{}
\author[2,3,4*]{Di Li}{}
\author[5]{Ashley K. Hyde}{}
\author[2]{Lei Qian}{}
\author[2]{Hualei Lyu}{}
\author[1]{Zhongzu Wu}{}


\address[{\rm1}]{College of Science/Department of Physics, Guizhou University, Guiyang 550025, China;}
\address[{\rm2}]{National Astronomical Observatories, Chinese Academy of Sciences, Beijing 100012, China;}
\address[{\rm3}]{Space Science Institute, Boulder, CO, USA;}
\address[{\rm4}]{Department of Astronomy, California Institute of Technology, CA, USA;}
\address[{\rm5}]{Astrophysics Group, Imperial College London, Blackett Laboratory, London, UK}
\maketitle \vspace{-3.5mm}
{\footnotesize\begin{center} Accepted February 27, 2014
\end{center}}\vspace*{-5mm}

\begin{center}
\rule{16.5cm}{0.4pt}
\parbox{16.5cm}
{\begin{abstract} Stars are born in dense cores of molecular clouds. The core mass function (CMF), which is the mass distribution of dense cores, is important for understanding the stellar initial mass function (IMF). We obtained 350 \micron\ dust continuum data using the SHARC-II camera at the Caltech Submillimeter Observatory (CSO) telescope. A 350 \micron\ map covering 0.25 ${deg}^{2}$ of the Ophiuchus molecular cloud was created by mosaicing 56 separate scans. The CSO telescope had an angular resolution of 9 \arcsec, corresponding to $1.2\times {10}^{3}\ $AU at the distance of the Ophiuchus molecular cloud (131 pc). The data was reduced using the Comprehensive Reduction Utility for SHARC-II (CRUSH). The flux density map was analyzed using the GaussClumps algorithm, within which 75 cores has been identified. We used the Spitzer c2d catalogs to separate the cores into 63 starless cores and 12 protostellar cores. By locating Jeans instabilities, 55 prestellar cores (a subcategory of starless cores) were also identified. The excitation temperatures, which were derived from  FCRAO ${}^{12}$CO data, help to improve the accuracy of the masses of the cores. We adopted a Monte Carlo approach to analyze the CMF with two types of functional forms; power law and log-normal. The whole and prestellar CMF are both well fitted by a log-normal distribution, with $\mu =-1.18\pm0.10,\ \sigma =0.58\pm0.05$ and $\mu =1.40\pm0.10,\ \sigma =0.50\pm0.05$ respectively. This finding suggests that turbulence influences the evolution of the Ophiuchus molecular cloud.
\end{abstract}}
\end{center}\vspace*{-0.6cm}

\begin{center}
\parbox{16.5cm}
{\bf\jiuhao ISM, molecular clouds, Ophiuchus, CMF}
\end{center}

\begin{center}

{}
\Cit{G. Y. Zhang, D. Li, et al. 350 \micron\ map of the Ophiuchus molecular cloud: core mass function. Sci China-Phys Mech Astron, in press.}
\end{center}

\wuhao\vspace*{1.5mm}

\begin{multicols}{2}

\renewcommand{\baselinestretch}{1.08} \baselineskip 12.2pt\parindent=10.8pt

\renewcommand{\thefootnote}

\section{Introduction}

\no Molecular cores are dense condensations within molecular clouds, in which stars are born [1-3]. The study of the CMF is therefore potentially important for understanding the$\ $IMF. The similarity between the CMF and the IMF suggests that$\ $stars may form at the same efficiency in all cores\ [4], although this remains unclear [5,6]. Protostellar cores have embedded self-luminous sources. Protostellar cores have lost some mass because of accretion onto the embedded protostar, or the extistence of bipolar outflows [7-9]. These changes could mean that protostellar cores no longer exhibit the same initial conditions as starless cores. Therefore, in

\vspace*{-3.0mm}
\noindent\rule{2.5cm}{0.4pt}\\[0.1mm]{\qihao *Corresponding author (email: ithaca.li@gmail.com)\\Accepted by SCIENCE CHINA  Physics, Mechanics \& Astronomy.}

\no order to better characterize the initial conditions of star formation, the differentiation of protostellar cores from starless cores needs to be investigated [10].

Herein we use both power law and log-normal functional form to fit the CMF. The power law form is the traditional form of the IMF, but in recent years it has been proposed that the low mass region of the CMF may be characterized by a log-normal function [10]. Ballesteros-Paredes et al. [11] calculated the CMF
using numerical models of turbulent fragmentation of molecular clouds, resulting in a form that does not follow a single power-law, and instead is more
similar to a log-normal function. Padoan and Nordlund [12] have deemed that the slope of the CMF depends on the slope of the turbulent energy spectrum. Previous research has shown that a log-normal distribution arises when the central limit theorem is applied to isothermal turbulence [13-15].

The aim of this study is to determine the CMF of the Ophiuchus molecular cloud, so as to advance the current understanding of how cores evolve into stars in a medium mass star forming region such as the Ophiuchus molecular cloud. We used core extraction and the CMF fitting techniques similar to those in Li et al. [16], which studied high mass quiescent cores in the Orion molecular cloud. The Ophiuchus molecular cloud is a medium mass star forming region [17]. There is no evidence for the existence of massive star clusters, H II regions and other usual marks of massive star formation, such as water masers, but there is one B star present within the cloud. There are several previously published dust continuum maps of the Ophiuchus molecular cloud. Motte et al. [18] mapped approximately 0.13 ${deg}^{2}$ at 1.3 mm, with an effective angular resolution of 15 \arcsec. The prestellar CMF can be fitted by a power law, $\rm dN/dM\sim {M}^{-1.5}$ for $\rm M <0.5$ \Ms, while the whole CMF appears to steepen to  $\rm dN/dM\sim {M}^{-2.5}$ for $\rm M >0.5$ \Ms. Johnstone et al. [19] mapped approximately 0.19 \squaredeg\ at 850 \micron, with an angular resolution of 15 \arcsec, identifying 55 cores. The CMF is fitted by two power laws, $\rm N(M)\sim {M}^{-\alpha }$, with $\alpha =1.0-1.5$ for $\rm M >0.6$ \Ms, and $\alpha =0.5$ for $\rm M \leq 0.6$ \Ms. Young et al. [20] mapped approximately 11 \squaredeg\ at 1.1 $mm$, with an angular resolution of 31 \arcsec, identifying 44 cores and a few candidates for cores. The CMF can be fitted by a power law function, $\rm N(M)\sim {M}^{-\alpha }$, with $\alpha =2.1\pm 0.3$ for $\rm M >0.5$ \Ms . Stanke et al. [21] mapped approximately 1.3 \squaredeg\ at 1.2 $mm$, with an angular resolution of 24 \arcsec, identifying 143 cores in this map, including 111 starless cores. The starless CMF is fitted by a broken power law, where the power law indices are -0.3, -1.3, and -2.3. Herein we use the data of dust continuum at 350 \micron, which is close to the peak of the dust spectral energy distribution, to map approximately 0.25 \squaredeg\ at this wavelengh. The CSO telescope provides 9 \arcsec angular resolution at 350 \micron\ which is 3 times better than Herschel [22]. Johnstone et al. [18] found 55 cores within a SCUBA/JCMT 850 \micron\ map of the Ophiuchus cloud. 54 cores reside within our 350 \micron\ map. They are indicated with the blue square symbol in Fig. 10. The cores unique to the 350 \micron\ map may have too high a temperature to have been detected by the Johnstone 850 \micron\ survey. Some of the cores apparently missing from this 350 \micron\ study in comparison with the 850 \micron\ study are likely to be cooler sources.
\section{Observations and data reduction}
\subsection{Observations}
\no We obtained data tracing the dust continuum in the Ophiuchus molecular cloud at 350\ \micron\ from the SHARC-II camera at the CSO telescope over three consecutive nights, 20th-22nd July 2009. SHARC-II [23] is mainly comprised of a 12 $\times$ 32 pixel array, covering a region of 2.59 \arcminute$\ \times\ $0.97 \arcminute, and having a beam size (FWHM) of 8.5 \arcsec at 350 \micron. We implemented the nonchopping box\_scan mode of SHARC-II, which is designed for scanning much wider areas than the scope of the SHARC-II array itself. The SHARC-II camera scanned the desired region quickly and repeatedly with a constant speed and uniform coverage [24]. Such redundancy enabled the sky image to be subsequently computed through software iteration [16]. Across the Ophiuchus molecular cloud region, we obtained a total of 56 separate scans in total in the format of FITS files, with four extensions: flux intensity, integration time, RMS and signal-to-noise. 56 separate scans were reduced using the Comprehensive Reduction Utility for SHARC-II (CRUSH) [25]. A large 350 \micron\ map (Fig. 7) of the Ophiuchus molecular cloud was created by mosaicing 56 separate scans of the Ophiuchus molecular cloud, with a field coverage of ~0.25 \squaredeg\ and an angular resolution of 9 \arcsec.  Fig. 1 contrasts our map with SCUBA/JCMT 850 \micron\ map $^{1)}$ of same region. and the improvement in angular resolution is apparent. 
\vspace{-2.0mm}
\begin{figure}[H]

\begin{minipage}{0.6\linewidth}
  \centerline{\includegraphics[scale=0.195]{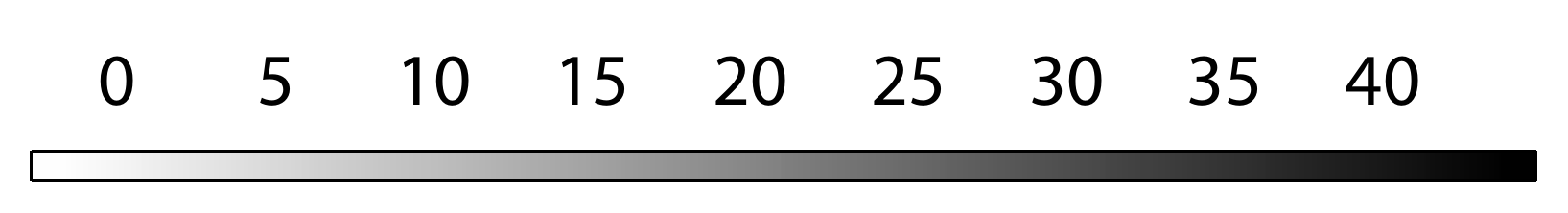}}
\end{minipage}
\hfill
\begin{minipage}{0.39\linewidth}
  \centerline{\includegraphics[scale=0.195]{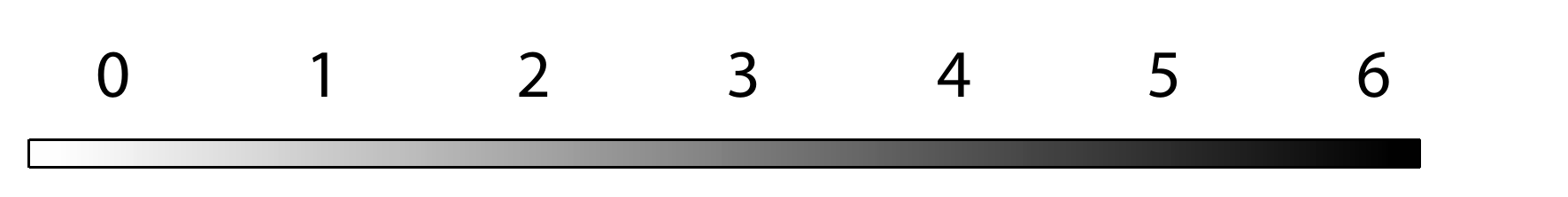}}
\end{minipage}
\vfill
\begin{minipage}{0.6\linewidth}
\vspace{-7.0mm}
  \centerline{\includegraphics[scale=0.288]{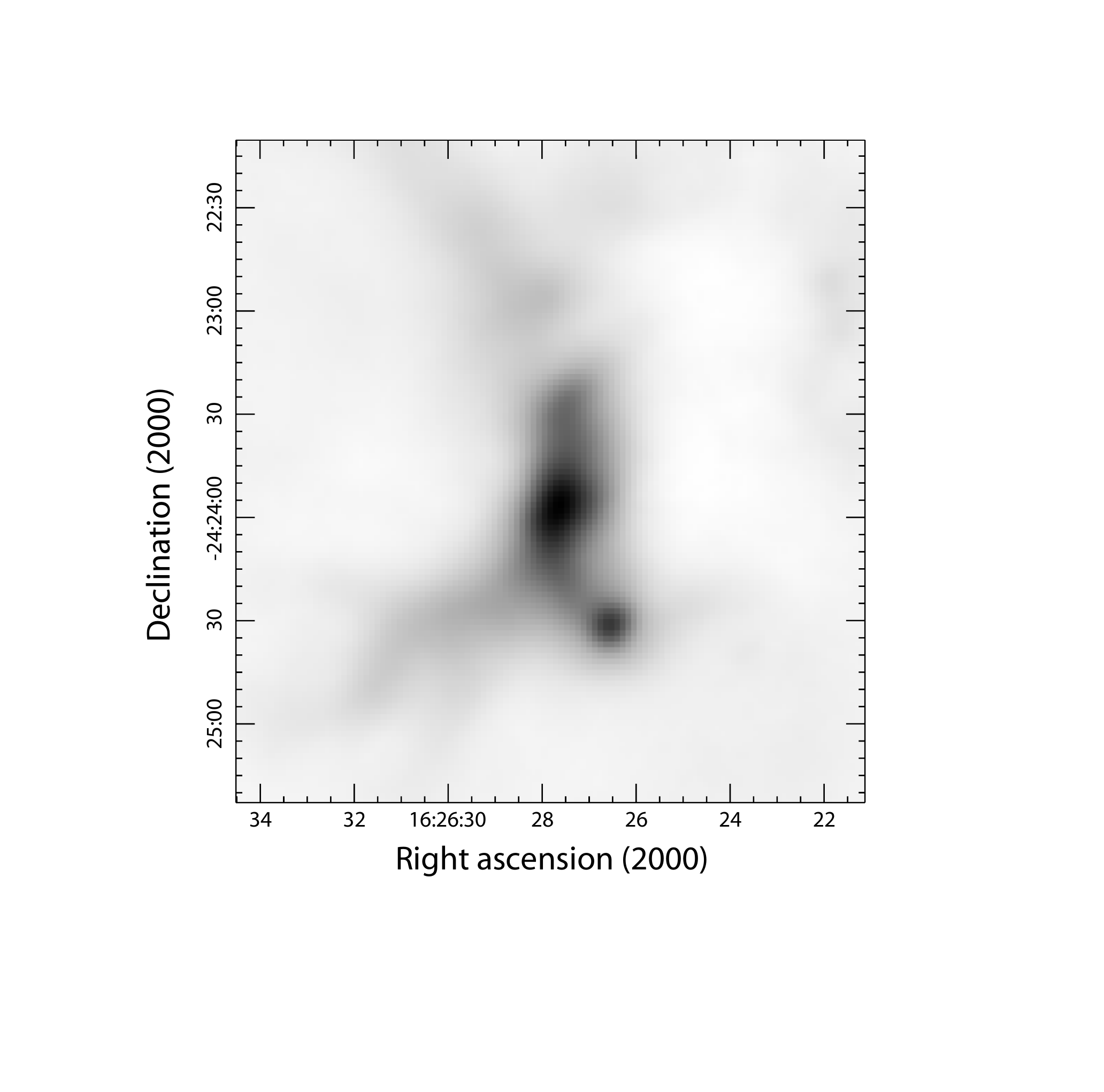}}
\end{minipage}
\hfill
\begin{minipage}{0.39\linewidth}
\vspace{-7.0mm}
\hspace*{-2.0mm}
  \centerline{\includegraphics[scale=0.288]{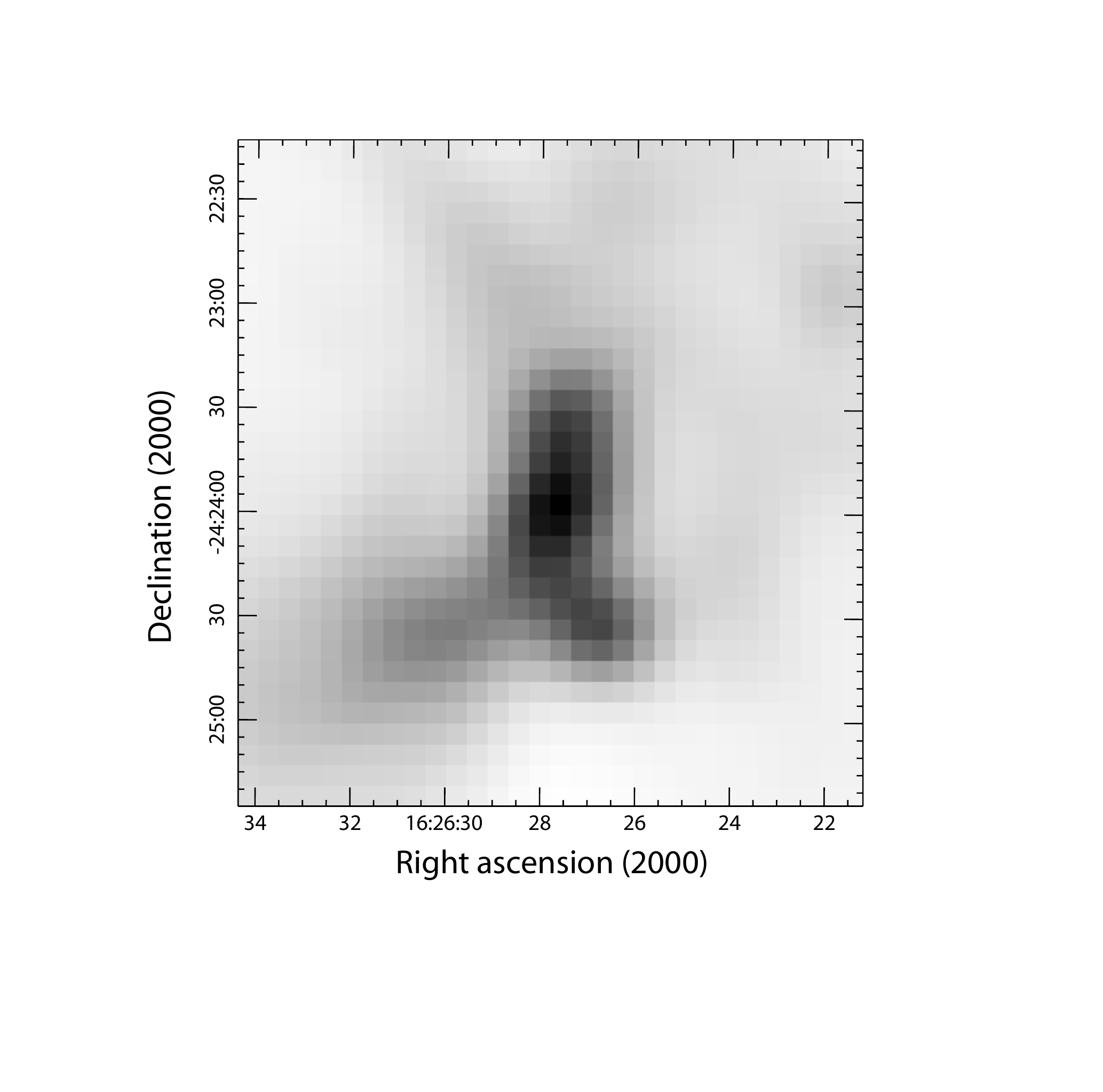}}
\end{minipage}
\vspace{-10.0mm}
\caption{The comparison between our 350 \micron\ map with SCUBA/JCMT 850 \micron\ map of same region. Left panel shows the map as imaged at 350 \micron\ in units of Jy per 8.5 \arcsec\ beam, with an angular resolution of 9 \arcsec. Sourcing of the map is in the lower right of this map, which does not end up being an identified core, because the FWHM value of the core is less than two beams, which is close to point-like. Right panel shows the map as imaged at 850 \micron\ in units of Jy per 15 \arcsec\ beam, with an angular resolution of 15 \arcsec.} 
\label{fig:example2}
\end{figure}
\subsection{Calibration}
The calibrators observed were Neptune on nights one and three and IRAS16293 on all three nights. IRAS16293 is a Young stellar object (YSO), which is close to our observational region. The calibrators were observed at approximate one-hour intervals with observational regions in between. The calibrator images seem to vary significantly from hour-to-hour because of the atmospheric effects. In order to reduce the impact of this variation, the science target data in these hour intervals was scaled by an appropriate factor. The corresponding scaling factor was found by dividing the
\end{multicols}
\vspace*{1mm}
\noindent\rule{2.5cm}{0.4pt}\\[0.1mm]
{\qihao{
\hspace*{3.5mm}1)COMPLETE team, "All Ophiuchus in Sub-millimeter continuum (850 microns)", http://hdl.handle.net/10904/10087 V2 [Version]\\
}}
\begin{multicols}{2}
corresponding flux of IRAS16293 by the last measured flux of $\ $IRAS16293 on the previous day. A calibration factor was required to determine how many of these detector flux units were equal to 1 Jy. We fitted a 2D Gaussian profile to the calibrator image, and from the peak flux and full width at half maximum (FWHM) fitted, the total flux of the object, 145.918 was calculated, in detector units. To find the calibration factor this was compared with the actual flux of Neptune on this date: 92.3424 Jy. Simple division showed that 1.58 detector units are equal to 1 Jy. Because IRAS16293 was found to be slightly elliptical in shape, only Neptune was analyzed in this way to ensure better reliability.

\subsection{Trimming the Map}
Certain scans near the outer regions of the map had a poor visual quality and appeared to suffer from a high level of noise. Through inspection of the integration time extension of the FITS file, it was found that these scans had a lower integration time relative to those more towards the center, approaching a
minimum value of ~2 seconds. Additionally, data at the extreme edge of the map had a particularly ‘grainy’ appearance. Since none of these areas contain any reliable data, they were trimmed. Using IDL, all pixels with an integration time of less than 2 seconds were recorded. The corresponding pixels in the flux intensity extension of the image were then set to a value of ‘NaN pixels’ (no finite value). These data were ignored automatically in the following process.

\section{Core Extraction}

Cores are assumed to have an elliptical shape. CUPID$^{2)}$, which mainly includes the subroutines FINDBACK and FINDCLUMPS, is a software package from Starlink [26]. We specified the GaussClumps algorithm in subroutine FINDCLUMPS, which is described by Stutzki and Gusten\ [27], to detect cores in the flux indensity map. The GaussClumps algorithm proceeds by fitting a Gaussian profile to the brightest peak in the data. It then subtracts the fit from the data and iterates, fitting a new ellipse to the brightest peak in the residuals.

\subsection{Sky background estimate}\vspace*{3mm}

The process we used to detect cores had two steps. First, we used FINDBACK to estimate the background. After that we used GaussClumps to measure the cores from the background-subtracted map. Estimating the background is the key to an accurate measurement. The fundamental principle is to wipe out large-scale diffuse material and let the cores stand out. There are negative depressions around the bright sources in our map, which must be removed before running FINDCLUMPS. The BOX value is the smoothing $\ $scale. Through experimenting with different BOX values and judging their visual effect on the map, we decided to set the BOX value to to 45 pixels, corresponding to 72.8 \arcsec. After checking the RMS dimension of our data, we estimated the RMS noise level. The RMS noise level of our no background-subtracted flux density map is about 0.11 Jy/beam. Fig. 7 is our background-subtracted map.


\subsection{Core Extraction}\vspace*{3mm}
After the background was removed, we used GaussClumps to find cores. Table~2 lists the parameters used in GaussClumps. For cores whose FWHM values have not been reduced to correct for the effect of the beam width, if the FWHM is equal to less than two beam widths then these cores are unconvincing and removed. For cores whose FWHM values have heen reduced to correct for the effect of the beam width, if they have a large axis ratio (major/minor$>$10), then these cores are unphysical [28] and are also removed. These selection criteria produced a final sample of 75 cores. We used GAIA to plot the colorized ellipsoidal shape of the cores on the flux density map (Fig. 8). Different color represents different temperature. We can define the radius of the core as;
\begin{equation}
R=\sqrt{{L}_{maj}*{L}_{min}}
\end{equation}
where $\rm {L}_{maj}$ is the major axis of the ellipse. $\rm {L}_{min}$ is the minor axis of the ellipse. The values of them are equal to FWHM of the fitted Gaussian. They are corrected to remove the effect of the instrumental beam width. We show the characteristics of the 75 cores in Table 1.

\subsection{Calculation of core mean temperatures}
Because 115 GHz (${}^{12}$CO J=1-0) is optically thick. The excitation temperatures derived from FCRAO\  ${}^{12}$CO data${}^{3)}$ can characterize the distribution of the gas temperature distribution. The dust and the gas couple together within high density regions of the molecular clouds, such as dense cores [29]. We therefore use the excitation temperatures in place of dust temperature $\rm {T}_{d}$ to estimate the mass. The FCRAO\ ${}^{12}$CO data is the uncorrected antenna temperature ($\rm {{T}_{A}}^{*}$) in Kelvin. To correct from $\rm {{T}_{A}}^{*}$ to the main beam temperature, we simply divide the $\rm {{T}_{A}}^{*}$ data by the main beam efficiency ${\eta }_{B}$. For 115 GHz (${}^{12}$CO J=1-0) , ${\eta }_{B}$ = 0.45. The excitation temperature is calculated by solving the following equation [30] as thus;
\begin{equation}
{T}_{max}=\frac{h\nu }{k}\left[ \frac{1}{{e}^{\frac{h\nu }{k{T}_{ex}}}-1}-\frac{1}{{e}^{\frac{h\nu }{k{T}_{bg}}}-1} \right]
\end{equation}

\no where h, k, $\nu$, $\rm {T}_{bg}$, $\rm {T}_{max}$, are Planck’s constant, Boltzmann’s constant, the central frequency of 115 GHz (${}^{12}$CO J=1-0), background temperature (assumed to be 2.7 K), and the spectral line’s peak value respectively. Using this method we obtained the excitation temperature map of the Ophiuchus molecular cloud. The mean value of the pixels within the FWHM of the core serve as the core mean temperatures. If a line is optically thick then we are only determining the temperature where the line became optically thick. Thus the cores, which may be more deeply embedded in the cloud, might actually have higher or lower temperatures, and this cannot be determined by this approach. There is a distribution of core mean temperatures (Fig. 2), reflecting the different nature and environment of each core.
\end{multicols}
\vspace*{1mm}
\noindent\rule{2.5cm}{0.4pt}\\[0.1mm]
{\qihao{
\hspace*{3.5mm}2)CUPID Users' Manual Version 0.0-38 http://starlink.jach.hawaii.edu/docs/sun255.htx/sun255.html\\
\hspace*{3.5mm}3)COMPLETE team, "FCRAO Ophiuchus 12CO cubes and map", http://hdl.handle.net/10904/10078 V2 [Version]
}}
\begin{multicols}{2}
\begin{figure}[H]
\begin{center}

\includegraphics[scale=0.48]{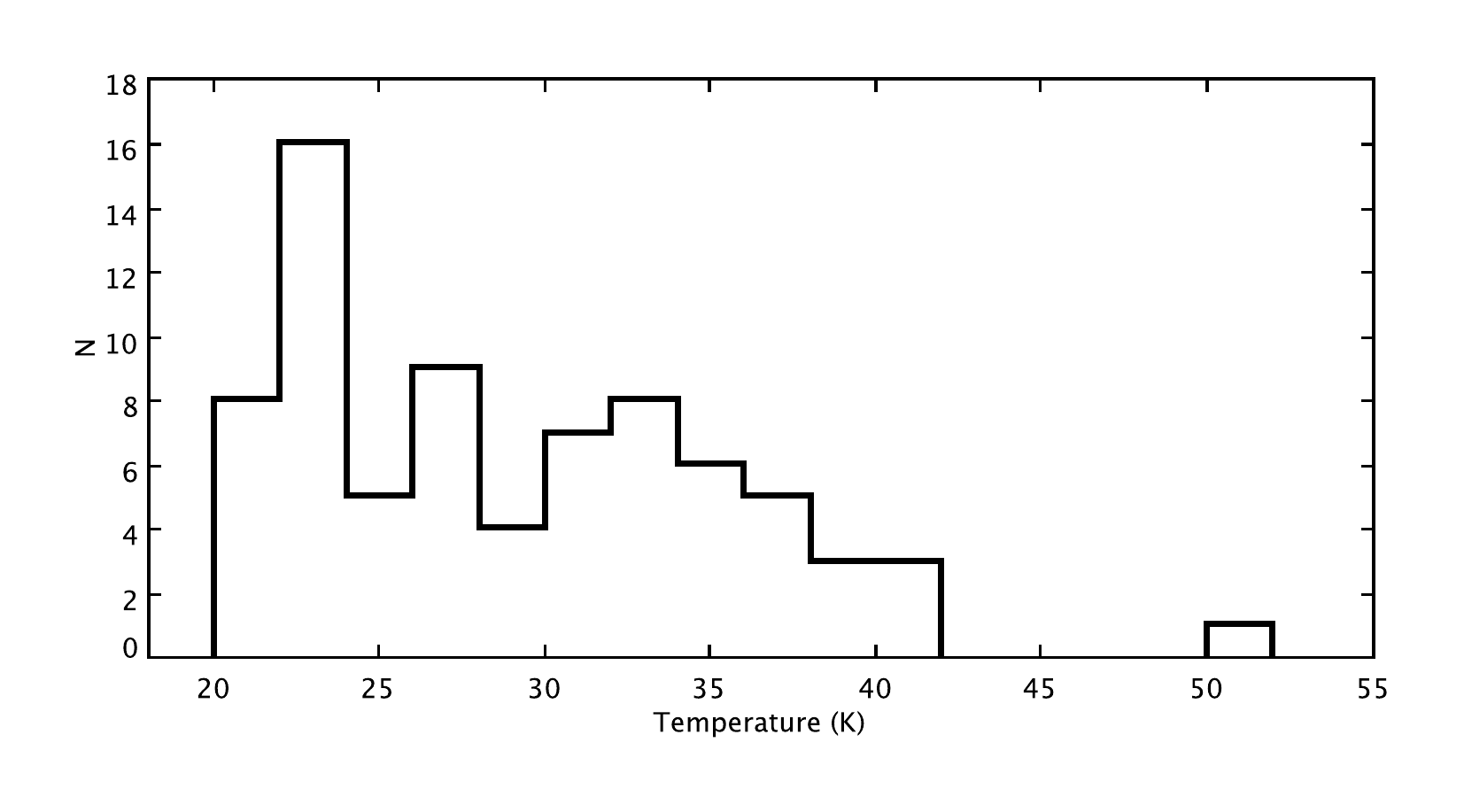}
\caption{Distribution of the mean temperatures of the cores.}
\end{center}
\end{figure}

\vspace{-10.0mm}

\subsection{Core masses estimates}
The main component of the molecular clouds is molecular hydrogen. Molecular hydrogen is a nonpolar molecule that is almost invisible at typical molecular cloud temperatures. The mass of a molecular cloud therefore needs to be inferred by observing other visible tracers. Continuum thermal emission from dust is often used as a tracer for the mass. The dust emission process is thermal, with dust grains emitting a modified blackbody spectrum. The total mass of each core given in Table 1, is estimated on the assumption that it is proportional to the measured 350 ${\mu m }$ flux, $\rm {F}_{350 \ \mu  m}$;

\begin{equation}
M=\frac{{D}^{2}{F}_{\nu }}{ B(\nu ,{T}_{d}){\kappa }_{\nu }},
\end{equation}
where D is the distance to the Ophiuchus molecular cloud, 131\ pc [31,32]. $\rm B(\nu ,{T}_{d})$ is the Planck function at dust temperature ${T}_{d}$. We use the excitation temperatures (see section 3.3) in place of dust temperature $\rm {T}_{d}$ to estimate the core total mass. $\rm {\kappa }_{350\ \mu  m}= 0.0101\ {cm}^{2} {g}^{-1}$ is the dust opacity. We assume that the dust emission at ${\kappa }_{\nu }$  is optically thin, and that ${\kappa }_{350 \ \mu  m}$ is independent of position within a core. The value of $\rm {\kappa }_{350 \ \mu  m}$ is found in Ossenkopf and Henning [33] Table 1, column (5) for dust grains with thin ice mantles, and a gas-to-dust mass ratio of 100. In Fig. 3, we display the most massive core Ophcso350-16. We took the low (20 K) and high (50 K) extremes from the excitation temperatures calculated and inferred the fractional change in mass between one extreme and the other (Fig. 4). It is clear that a change in the core temperature induces a significant change in the estimated core mass.
\vspace{-15.0mm}
\begin{figure}[H]
\begin{center}
\hspace*{-5mm}
\includegraphics[scale=0.48,angle=0]{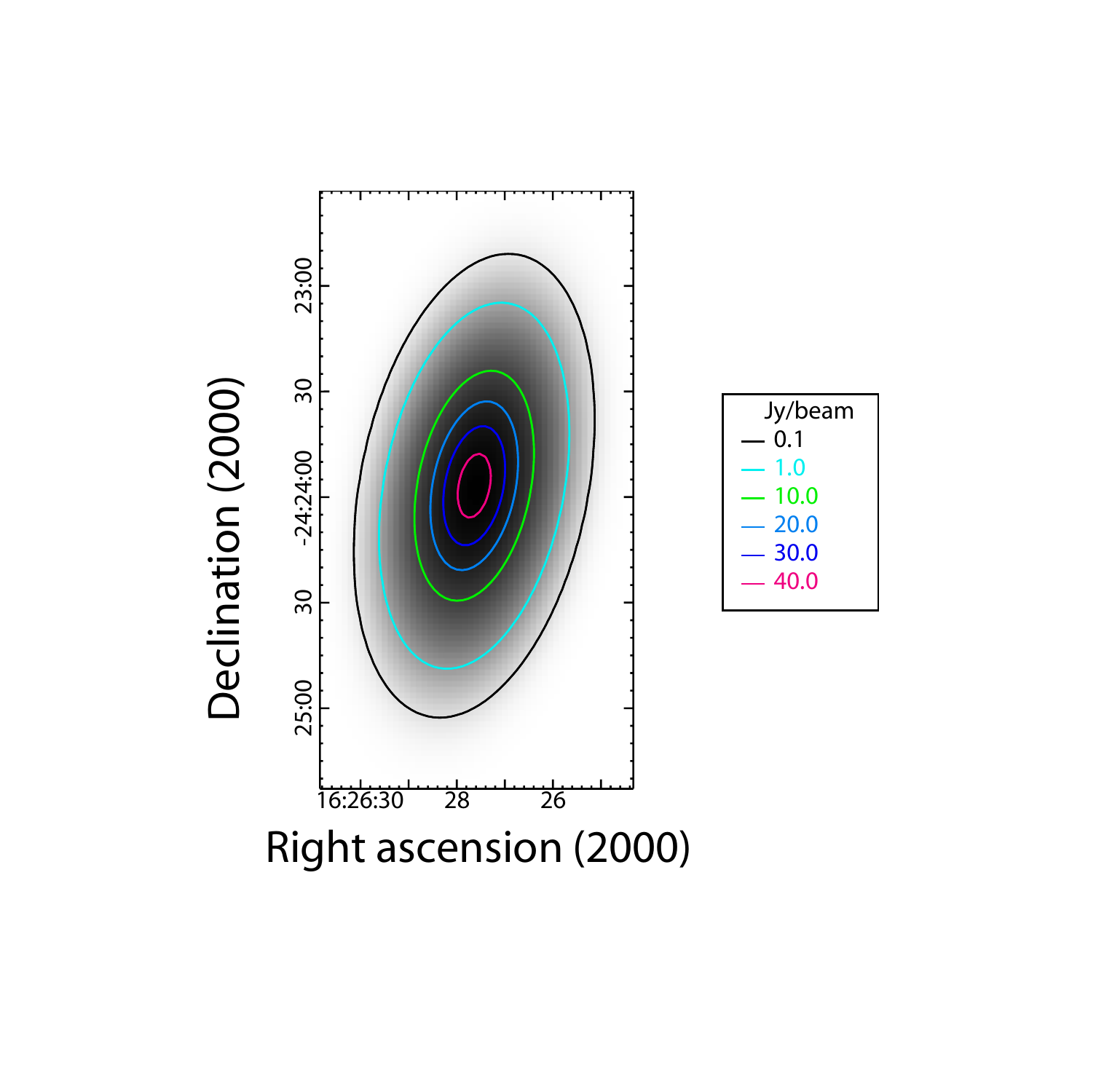}
\vspace{-7.0mm}
\caption{Most massive core: Ophcso350-16. Image above is plotted in units of Jy/beam. Mass of this core is 9.86 M$_\odot$. Peak of this core is 49.3 Jy/beam, as well as the mean temperature at 39.4K. }
\end{center}
\end{figure}
\begin{figure}[H]
\begin{center}
\hspace*{0mm}
\includegraphics[scale=0.48]{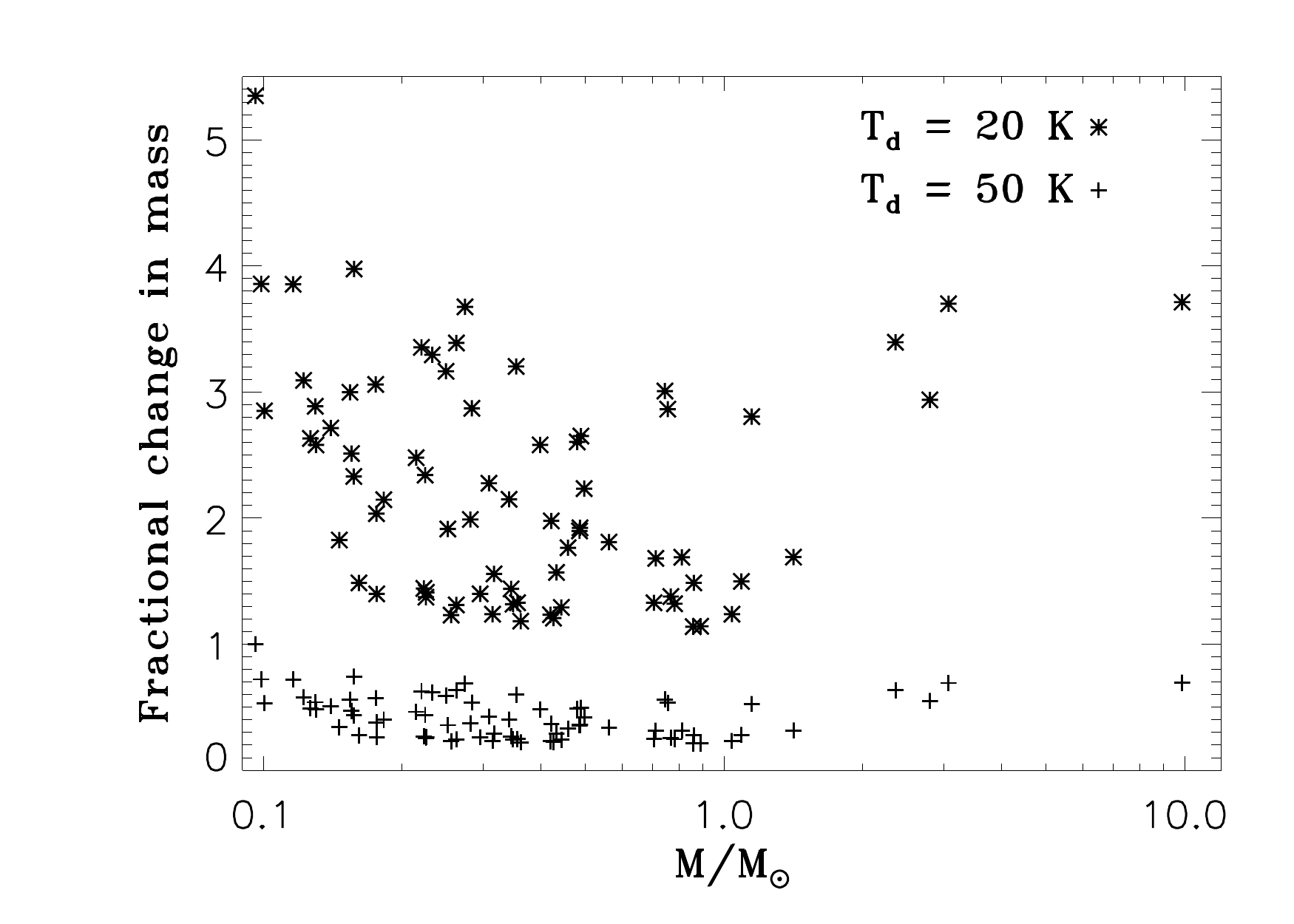}
\caption{Fractional change in mass. Assuming $\rm {T}_{d}=20\ K$ and $\rm {T}_{d}=50\ K$, we estimated core masses using equation Eq. (3). $\rm {M}_{20\ K}/M$ are marked with asterisks, the mean value of which is 2.25. $\rm {M}_{50\ K}/M$ are marked with crosses, the mean value of which is 0.42.}
\end{center}
\end{figure}

\subsection{Distinguishing the types of cores}

A considerable amount of research has been done on separating starless core populations from protostellar core populations [34,35], by setting up appropriate constraints according to the position of the YSO with respect to the cores, in infrared colors or magnitudes. In our work starless and protostellar cores are differentiated using the Spitzer c2d surveys, by identifying infrared sources that may be associated with a given core. We used Gator${}^{4)}$ to select c2d candidate YSO clouds catalog of the Ophiuchus molecular cloud [36]. A sampling of 147 YSOs are indicated with a red cross symbol in Fig. 9. If the distance between the center of the core and the position of the YSO is less than the core’s radius, the cores are assumed to be protostellar cores. Using this method, we found 12 protostellar cores and we deemed the remaining 63 cores to be starless cores. Prestellar cores are a subcategory of starless cores, and are dense, gravitationally bound clumps of gas within a molecular cloud. Prestellar cores are characterised by a Jeans instability. For a critical Bonnor-Ebert sphere the Jeans Length [37] is;
\begin{equation}
{\lambda }_{J}\approx 0.763{\left( \frac{{k}_{B}T}{G\rho \mu} \right)}^{1/2}
\end{equation}
where ${k}_{B}$ is Boltzmann's constant, G is the Gravitational Constant, T is the mean temperature of the core, $\rho$\ is the is the mean density of the core, $\mu$ is the mass of one molecule (taking a mean molecular weight $\rm \mu =2.3\ g\ /\ {N}_{A}$, where $\rm {N}_{A}$ is the Avogadro constant ). This will collapse if the radius is smaller than the Jeans length. Under this condition, a core can be considered to be prestellar core. Using this method, we found 55 prestellar cores. Sadavoy et al. [38] also investigated for cores with Jeans instability in Ophiuchus and many other Gould Belt clouds.

\begin{figure}[H]
\begin{center}
\hspace*{0mm}
\includegraphics[scale=0.48,angle=0]{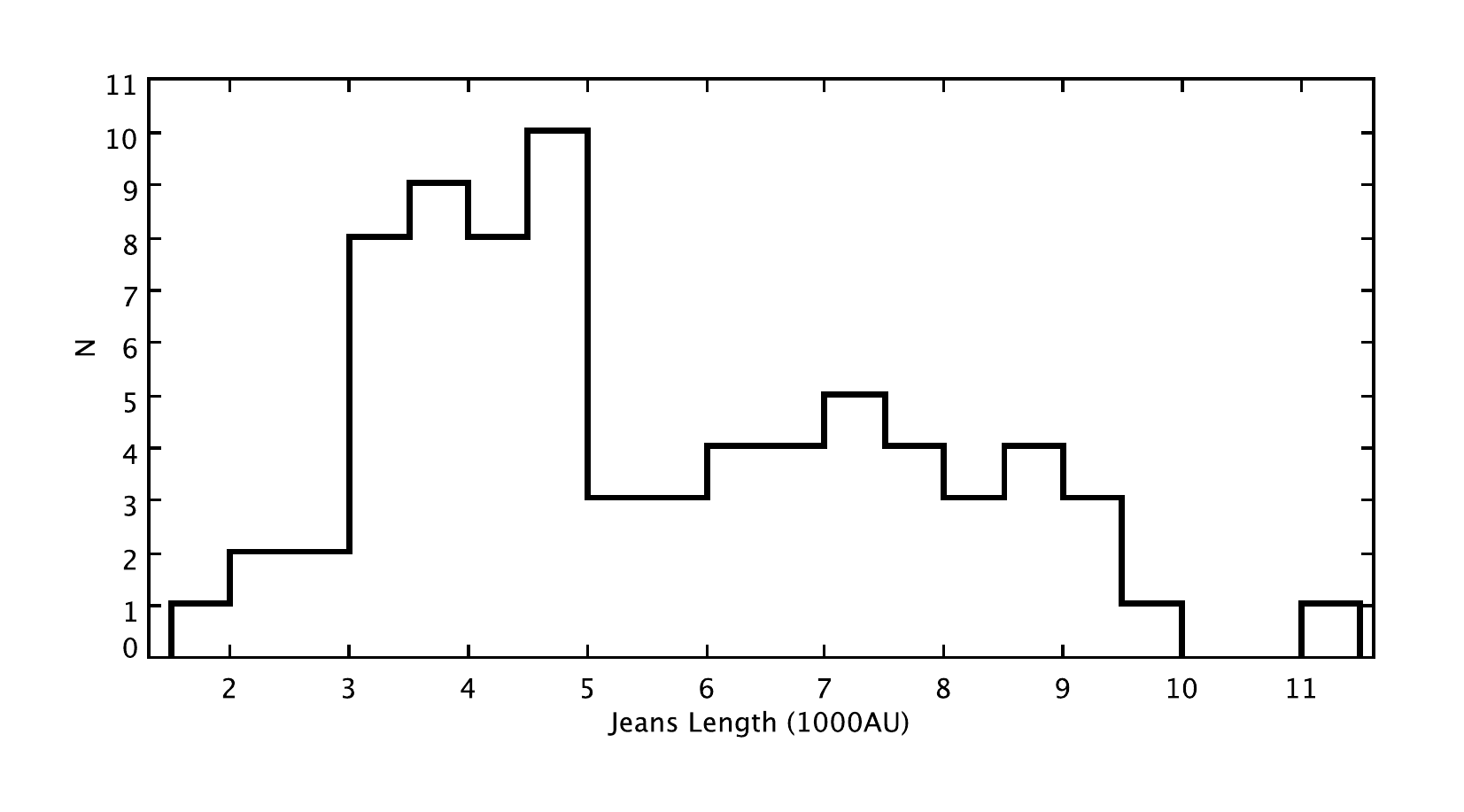}
\caption{Distribution of cores' Jeans Length. For a critical Bonnor-Ebert sphere its Jeans Length is calculated by the Eq. (4).}
\end{center}
\end{figure}

\begin{figure}[H]
\centering
\includegraphics[scale=0.48]{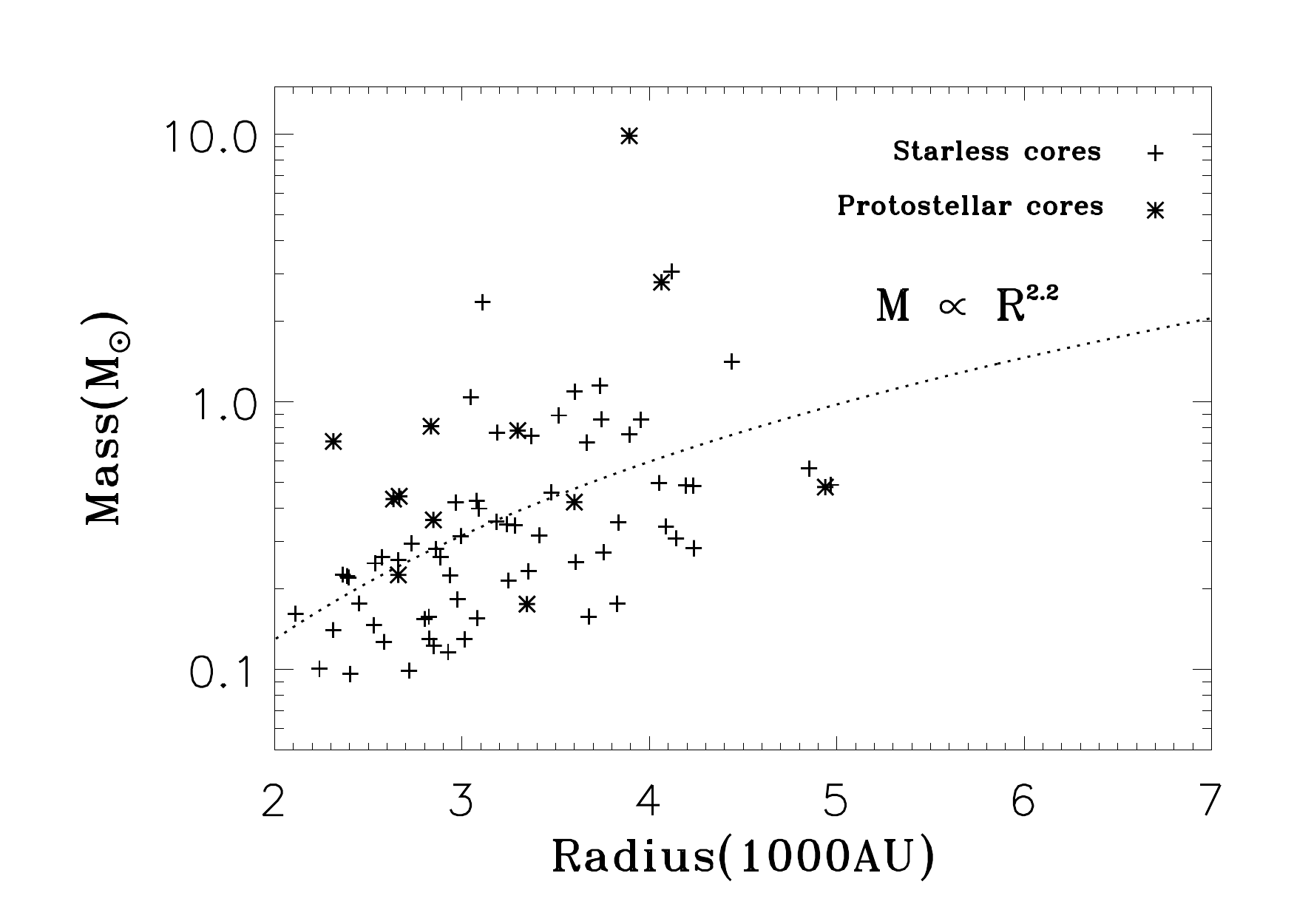}
\caption{Starless and protostellar cores plotted in terms of radius against mass. Starless cores are marked with crosses. Protostellar cores are marked with asterisks. The mass$-$radius relationship of the cores can be fitted with a power law $\rm M\propto {R}^{2.2}$.}
\end{figure}

\subsection{Minimum detection mass}\vspace*{3mm}
Whether the cores can be detected is limited by the angular resolution of the telescope and the physical properties of the cores; particularly size and flux density. According to our experience, the objects would most likely be taken as point sources if they are too close to the telescope beam size. Anything smaller than the telescope beam is likely to be noise, since even a point source in the sky might look like the telescope beam in the map. For a point source, if the minimum flux density is 5 $\sigma $ ($\sigma $: the global root mean square (RMS) noise level in the data), in one beam, we can locate sources at considerably significant confidence levels. If the source occupies N beams, a lower flux density in each pixel is adequate. To achieve the same statistical significance as that of the detection of the point source, the required flux density in each pixel is scaled as $\sqrt{\rm N}$. According to Eq. (3), for an isothermal, optically thin dust core, its flux is proportional to its mass. We calculated the minimum detectable mass for a core of certain size [16]:
\begin{equation}
{M}_{det}={M}_{point}\sqrt{N},
\end{equation}
where M${}_{point}$ is the minimum mass of a point source. We define the flux of a point source as 5 $\sigma $, that is, 0.2 Jy. The temperature is set to 30 K, which is the RMS of the mean temperatures the cores. The M${}_{point}$ calculated using the Eq. (3) is 0.005 M${}_{\odot}$. The upper mass boundary of the region of incompleteness, 0.09 M${}_{\odot}$, is set by a core of $2.0\times {10}^{4}\ $Au corresponding to an angular size of approximately 2.6 \arcminute. This limit corresponds to the largest linear dimension of the SHARC II array. The Ophiuchus molecular cloud is complete above the upper mass boundary. The red lines in Fig. 10 represent the upper mass boundary.
\section{Core mass function}
The exact shape of the local IMF is still somewhat uncertain [39]. Salpeter [40] initially postulated a power law form for the IMF:

\begin{equation}
\frac{dN}{dM}\propto {M}^{\alpha},\alpha= -2.35, \:  0.4{M}_{_\odot}<M<10{M}_{_\odot},
\end{equation}

where N is the number of stars with mass M, and $\alpha$ is a dimensionless exponent. Miller and Scalo [41] suggested that the exponent ($\alpha $) of the IMF approached zero below 1 M$_\odot$. Kroupa [42], extended the IMF of Salpeter [40] into a four-part power law.
When $\alpha$ is not equal to -1, the cumulative form of Eq. (6) is;
\begin{equation}
N(>{M}_{min})=\int_{{M}_{min}}^{{M}_{max}}dN\sim \frac{1}{\alpha +1}{{M}}^{\alpha +1}_{max}-\frac{1}{\alpha +1}{{M}}^{\alpha +1}_{min},
\end{equation}

where $\rm {M}_{max}$ is the maximum mass for a certain sample, and $\rm {M}_{min}$ is the minimum mass.
It is readily apparent that Eq. (7) is also a power law, but when $\alpha$ is equal to -1, the cumulative form of Eq. (6) becomes a log function [16].\\
Chabrier [43] suggests a log-normal form of the IMF. Some additional observations also show evidence for a log-normal form for the CMF [10, 21] such that;

\begin{equation}
\frac{dN}{d{\rm log}M}\propto {\rm exp}\left[ -\frac{{\left({\rm log}M-\mu  \right)}^{2}}{2{\sigma }^{2}}\right],
\end{equation}
where, $\mu$ and ${\sigma }^{}$, denote respectively the mean mass and the variance in units of $\rm logM$ [44].
The cumulative form of Eq. (8) is the complementary error function;
\begin{equation}
N(>M)=1-erf\left( \frac{logM-\mu }{\sqrt{2}\sigma } \right).
\end{equation}
We used 75 cores to analyze the CMF. The power law cumulative functional form (Eq. (7)) and log-normal cumulative functional form (Eq. (9)) were used to fit the CMF. In order to avoid introducing an error in the direct fitting of the power law cumulative functional form of the CMF, we used the Monte Carlo method adopted by Qian et al. [30], which

\end{multicols}
\vspace*{1mm}
\noindent\rule{2.5cm}{0.4pt}\\[0.1mm]
{\qihao{
\hspace*{3.5mm}4)http://irsa.ipac.caltech.edu/applications/Gator/}}
\begin{multicols}{2}

\end{multicols}
\begin{table}[p]
\label{tab:table1}
\caption{Properties of the cores \hspace*{3mm}}
\begin{center}\footnotesize \doublerulesep 0.15pt \tabcolsep 2.7pt
\begin{tabular*}{\textwidth}{ccccccccccccccc}\toprule[0.65pt]  
Index & $\rm {RA}^{a}$ & $\rm {DEC}^{a}$  & ${{L}_{maj}}^{b}$ &  ${{L}_{min}}^{b}$  & ${\theta}^{c}$ & $\rm {R}^{d}$ &${{\lambda }_{J}}^{e}$ &  $\rm {P}^{f}$  & $\rm {F}^{g}$ &  ${\rho}^{h}$  & ${n}^{h}$ & ${T}^{i}$ & ${M}^{j}$ &${type}^{k}$  \\
     & (\rm J2000) & \rm \hspace{0.3cm}(J2000) &(\arcsec)& (\arcsec)& ($\ {}^{\circ}\ $) & (${10}^{3}$AU) & (${10}^{3}$AU) & (\rm Jy/beam) & (Jy) & $({10}^{-18}g/{cm}^{3})$ & $({10}^{5}/{cm}^{3})$ & $ (K) $ &(M$_\odot$) & \\
\hline
 Ophcso350-1 &  16h26m 2.2s   &  -24d32m15.9s &  20.2 &  16.6 & 120.5 &   2.4 &   8.5 &   1.8 &   8.6 &   1.0 &   2.6 &  50.0 &  0.10 & SP \\
 Ophcso350-2 &  16h26m 7.6s   &  -24d20m27.3s &  41.2 &  19.9 &  47.6 &   3.8 &   8.7 &   1.4 &  16.7 &   0.7 &   1.9 &  39.2 &  0.27 & SP \\
 Ophcso350-3 &  16h26m10.1s   &  -24d19m44.1s &  30.0 &  21.9 &   6.2 &   3.4 &   7.7 &   1.3 &  12.8 &   0.9 &   2.3 &  36.7 &  0.23 & SP \\
 Ophcso350-4 &  16h26m10.1s   &  -24d23m13.2s &  24.9 &  15.0 &  45.7 &   2.5 &   4.8 &   2.4 &  13.1 &   2.2 &   5.6 &  35.8 &  0.25 & SP \\
 Ophcso350-5 &  16h26m12.3s   &  -24d37m16.6s &  18.4 &  17.0 & 163.5 &   2.3 &   5.4 &   1.4 &   6.3 &   1.6 &   4.2 &  32.7 &  0.14 & SP \\
 Ophcso350-6 &  16h26m14.4s   &  -24d22m45.2s &  25.0 &  21.1 &  15.8 &   3.0 &   8.4 &   0.8 &   6.2 &   0.7 &   1.8 &  33.9 &  0.13 & SP \\
 Ophcso350-7 &  16h26m14.8s   &  -24d25m 6.5s &  31.6 &  24.9 & 168.5 &   3.7 &  11.4 &   0.9 &  10.4 &   0.4 &   1.2 &  41.2 &  0.16 & SP \\
 Ophcso350-8 &  16h26m18.0s   &  -24d37m16.0s &  26.8 &  14.5 &  75.1 &   2.6 &   6.6 &   1.0 &   5.5 &   1.0 &   2.7 &  32.2 &  0.13 & SP \\
 Ophcso350-9 &  16h26m18.8s   &  -24d25m10.5s &  29.5 &  14.6 & 149.7 &   2.7 &   9.0 &   1.0 &   6.3 &   0.7 &   1.8 &  40.4 &  0.10 & SP \\
Ophcso350-10 &  16h26m21.4s   &  -24d23m36.0s &  24.9 &  15.5 & 134.1 &   2.6 &   4.9 &   2.7 &  14.8 &   2.2 &   5.7 &  37.3 &  0.26 & SP \\
Ophcso350-11 &  16h26m21.9s   &  -24d19m34.3s &  32.3 &  14.4 & 139.1 &   2.8 &   6.5 &   0.9 &   6.1 &   1.0 &   2.6 &  30.1 &  0.16 & SP \\
Ophcso350-12 &  16h26m22.6s   &  -24d25m 8.9s &  28.9 &  17.3 & 153.5 &   2.9 &   9.3 &   1.0 &   7.4 &   0.7 &   1.7 &  40.4 &  0.12 & SP \\
Ophcso350-13 &  16h26m23.7s   &  -24d20m35.0s &  70.3 &  20.2 &  91.3 &   4.9 &   8.9 &   1.0 &  20.8 &   0.6 &   1.5 &  32.0 &  0.48 & PR \\
Ophcso350-14 &  16h26m24.3s   &  -24d21m52.8s &  41.7 &  21.2 &  56.0 &   3.9 &   5.1 &   2.8 &  35.9 &   1.8 &   4.7 &  33.8 &  0.75 & SP \\
Ophcso350-15 &  16h26m26.4s   &  -24d22m21.8s &  27.9 &  23.7 &  39.7 &   3.4 &   4.2 &   4.0 &  37.2 &   2.7 &   7.2 &  34.7 &  0.74 & SP \\
Ophcso350-16 &  16h26m27.5s   &  -24d23m56.8s &  44.1 &  20.0 & 113.0 &   3.9 &   1.5 &  49.3 & 609.2 &  23.6 &  61.8 &  39.4 &  9.86 & PR \\
Ophcso350-17 &  16h26m27.6s   &  -24d23m21.2s &  27.0 &  20.9 & 176.7 &   3.1 &   2.2 &  17.0 & 133.0 &  11.1 &  29.0 &  37.3 &  2.35 & SL \\
Ophcso350-18 &  16h26m28.0s   &  -24d26m29.6s &  27.0 &  24.2 &  87.9 &   3.3 &   8.6 &   0.9 &   8.9 &   0.7 &   1.7 &  35.1 &  0.18 & PR \\
Ophcso350-19 &  16h26m28.5s   &  -24d22m49.8s &  36.1 &  26.7 &  48.0 &   4.1 &   2.9 &  10.2 & 136.7 &   5.9 &  15.4 &  34.3 &  2.80 & PR \\
Ophcso350-20 &  16h26m29.5s   &  -24d21m27.2s &  20.0 &  14.6 & 169.3 &   2.2 &   6.1 &   1.1 &   4.8 &   1.3 &   3.3 &  33.7 &  0.10 & SP \\
Ophcso350-21 &  16h26m29.9s   &  -24d24m29.5s &  38.2 &  25.9 &  38.9 &   4.1 &   3.0 &  13.6 & 189.0 &   6.2 &  16.2 &  39.4 &  3.07 & SL \\
Ophcso350-22 &  16h26m30.2s   &  -24d22m19.2s &  36.2 &  22.5 & 136.4 &   3.7 &   3.9 &   4.6 &  53.6 &   3.1 &   8.1 &  33.4 &  1.15 & SP \\
Ophcso350-23 &  16h26m31.4s   &  -24d21m53.1s &  26.4 &  21.1 &  80.5 &   3.1 &   4.8 &   2.2 &  17.1 &   1.9 &   5.0 &  31.8 &  0.40 & SP \\
Ophcso350-24 &  16h26m32.9s   &  -24d26m15.8s &  34.3 &  25.0 &  90.4 &   3.8 &   7.6 &   1.5 &  18.8 &   0.9 &   2.3 &  36.0 &  0.35 & SP \\
Ophcso350-25 &  16h26m33.2s   &  -24d21m27.8s &  33.3 &  18.4 & 122.4 &   3.2 &   7.0 &   1.0 &   8.9 &   0.9 &   2.3 &  31.1 &  0.21 & SP \\
Ophcso350-26 &  16h26m33.4s   &  -24d24m56.6s &  19.1 &  17.4 & 163.4 &   2.4 &   4.8 &   2.6 &  12.3 &   2.3 &   5.9 &  37.1 &  0.22 & SP \\
Ophcso350-27 &  16h26m35.3s   &  -24d26m31.2s &  24.6 &  19.2 & 142.2 &   2.8 &   8.2 &   0.9 &   6.3 &   0.7 &   2.0 &  35.3 &  0.12 & SP \\
Ophcso350-28 &  16h26m35.9s   &  -24d17m54.2s &  28.9 &  26.2 &  60.4 &   3.6 &   7.1 &   0.7 &   8.0 &   0.8 &   2.0 &  27.1 &  0.25 & SP \\
Ophcso350-29 &  16h26m36.4s   &  -24d24m35.9s &  48.2 &  21.7 &  38.2 &   4.2 &   9.5 &   0.9 &  13.5 &   0.5 &   1.4 &  33.8 &  0.28 & SP \\
Ophcso350-30 &  16h26m36.5s   &  -24d20m31.5s &  50.1 &  19.4 & 154.2 &   4.1 &   7.6 &   0.8 &  12.2 &   0.7 &   1.8 &  28.8 &  0.34 & SP \\
Ophcso350-31 &  16h26m39.7s   &  -24d24m 6.0s &  56.1 &  17.8 &  13.4 &   4.1 &   8.2 &   0.8 &  11.7 &   0.6 &   1.6 &  29.7 &  0.31 & SP \\
Ophcso350-32 &  16h26m40.0s   &  -24d14m56.6s &  22.7 &  15.4 &  71.3 &   2.5 &   4.4 &   0.8 &   4.1 &   1.7 &   4.4 &  23.2 &  0.18 & SP \\
Ophcso350-33 &  16h26m42.8s   &  -24d26m12.1s &  27.9 &  16.4 & 157.0 &   2.8 &   7.0 &   1.1 &   7.7 &   1.0 &   2.6 &  34.7 &  0.15 & SP \\
Ophcso350-34 &  16h26m43.5s   &  -24d17m26.2s &  34.9 &  21.7 &  91.4 &   3.6 &   3.2 &   2.5 &  27.2 &   3.3 &   8.6 &  24.0 &  1.09 & SL \\
Ophcso350-35 &  16h26m47.9s   &  -24d23m52.7s &  26.3 &  19.6 &  45.5 &   3.0 &   6.4 &   0.9 &   6.5 &   1.0 &   2.6 &  28.8 &  0.18 & SP \\
Ophcso350-36 &  16h26m54.8s   &  -24d14m55.5s &  28.1 &  24.2 &  13.8 &   3.4 &   5.5 &   0.8 &   8.2 &   1.1 &   2.9 &  24.4 &  0.32 & SP \\
Ophcso350-37 &  16h26m56.6s   &  -24d36m38.4s &  57.0 &  24.1 &   9.0 &   4.9 &   7.3 &   0.8 &  16.9 &   0.7 &   1.8 &  26.3 &  0.56 & SP \\
Ophcso350-38 &  16h26m57.9s   &  -24d34m22.6s &  40.5 &  25.8 &  24.0 &   4.2 &   6.5 &   1.0 &  15.3 &   0.9 &   2.4 &  27.0 &  0.49 & SP \\
Ophcso350-39 &  16h26m58.2s   &  -24d37m52.1s &  22.3 &  18.5 &  86.5 &   2.7 &   4.4 &   0.9 &   5.1 &   1.7 &   4.4 &  23.0 &  0.23 & PR \\
Ophcso350-40 &  16h26m58.9s   &  -24d34m54.3s &  20.7 &  18.0 &   6.2 &   2.5 &   5.4 &   0.8 &   4.4 &   1.3 &   3.3 &  26.5 &  0.15 & SP \\
Ophcso350-41 &  16h27m 2.5s   &  -24d38m49.3s &  28.5 &  20.7 & 135.4 &   3.2 &   4.5 &   0.9 &   7.9 &   1.6 &   4.1 &  22.7 &  0.36 & SP \\
Ophcso350-42 &  16h27m 4.5s   &  -24d39m14.4s &  35.0 &  22.4 &  42.6 &   3.7 &   4.0 &   1.4 &  15.6 &   2.0 &   5.3 &  22.7 &  0.70 & SP \\
Ophcso350-43 &  16h27m 5.2s   &  -24d36m30.2s &  25.0 &  16.5 & 145.0 &   2.7 &   3.1 &   1.6 &   9.5 &   3.3 &   8.7 &  22.4 &  0.44 & PR \\
Ophcso350-44 &  16h27m11.1s   &  -24d27m35.9s &  61.6 &  23.4 & 105.5 &   5.0 &   9.0 &   1.0 &  21.6 &   0.6 &   1.5 &  32.3 &  0.49 & SP \\
Ophcso350-45 &  16h27m11.5s   &  -24d29m24.7s &  47.6 &  21.6 & 164.0 &   4.2 &   6.4 &   1.0 &  15.6 &   0.9 &   2.4 &  27.2 &  0.49 & SP \\
Ophcso350-46 &  16h27m11.7s   &  -24d25m58.5s &  29.7 &  15.7 & 132.3 &   2.8 &   7.4 &   0.8 &   5.6 &   0.8 &   2.1 &  31.8 &  0.13 & SP \\
Ophcso350-47 &  16h27m11.8s   &  -24d37m58.3s &  30.2 &  18.3 &  97.2 &   3.1 &   3.8 &   1.1 &   8.6 &   2.1 &   5.4 &  21.7 &  0.43 & SP \\
Ophcso350-48 &  16h27m15.2s   &  -24d16m31.9s &  17.2 &  15.1 &  69.4 &   2.1 &   3.7 &   1.1 &   4.0 &   2.4 &   6.3 &  23.9 &  0.16 & SP \\
Ophcso350-49 &  16h27m15.4s   &  -24d30m41.3s &  37.7 &  20.0 & 135.9 &   3.6 &   5.5 &   1.3 &  13.9 &   1.3 &   3.3 &  27.6 &  0.42 & PR \\
Ophcso350-50 &  16h27m16.6s   &  -24d40m44.1s &  28.8 &  18.1 &  18.1 &   3.0 &   4.3 &   0.9 &   6.5 &   1.7 &   4.3 &  22.0 &  0.31 & SP \\
Ophcso350-51 &  16h27m16.8s   &  -24d27m42.1s &  26.6 &  20.8 &  86.3 &   3.1 &   7.7 &   0.8 &   6.5 &   0.8 &   2.0 &  31.3 &  0.16 & SP \\
Ophcso350-52 &  16h27m17.8s   &  -24d28m54.6s &  21.7 &  18.6 &  18.0 &   2.6 &   3.2 &   2.0 &  11.3 &   3.3 &   8.7 &  24.5 &  0.43 & PR \\
Ophcso350-53 &  16h27m19.0s   &  -24d39m41.1s &  35.1 &  14.6 &  75.6 &   3.0 &   3.7 &   1.1 &   8.6 &   2.3 &   5.9 &  21.9 &  0.42 & SP \\
Ophcso350-54 &  16h27m19.7s   &  -24d27m10.0s &  46.8 &  20.4 & 145.9 &   4.1 &   6.3 &   1.3 &  18.5 &   1.1 &   2.8 &  29.4 &  0.50 & SP \\
Ophcso350-55 &  16h27m20.8s   &  -24d14m34.7s &  30.9 &  16.3 &   9.6 &   2.9 &   5.8 &   1.2 &   8.7 &   1.3 &   3.3 &  30.1 &  0.22 & SP \\
Ophcso350-56 &  16h27m21.6s   &  -24d39m49.5s &  33.9 &  25.2 &  98.5 &   3.8 &   3.4 &   1.8 &  21.4 &   2.6 &   6.8 &  22.0 &  1.04 & SL \\
Ophcso350-57 &  16h27m21.6s   &  -24d24m50.1s &  28.2 &  19.2 & 144.1 &   3.0 &   6.7 &   0.8 &   6.0 &   0.9 &   2.3 &  28.0 &  0.18 & SP \\

\bottomrule[0.65pt] 
\end{tabular*}
\end{center}
\end{table}
\begin{table}[p]
\label{tab:table1}
{\bf\bahao Table 1 }{\bahao (continued)}
\begin{center}\footnotesize \doublerulesep 0.15pt \tabcolsep 2.7pt
\begin{tabular*}{\textwidth}{ccccccccccccccc}\toprule[0.65pt]  
Index & $\rm {RA}^{a}$ & $\rm {DEC}^{a}$  & ${{L}_{maj}}^{b}$ &  ${{L}_{min}}^{b}$  & ${\theta}^{c}$ & $\rm {R}^{d}$ &${{\lambda }_{J}}^{e}$ &  $\rm {P}^{f}$  & $\rm {S}^{g}$ &  ${\rho}^{h}$  & ${n}^{h}$ & ${T}^{i}$ & ${M}^{j}$ &${type}^{k}$  \\
     & (\rm J2000) & \rm \hspace{0.3cm}(J2000) &(\arcsec)& (\arcsec)& ($\ {}^{\circ}\ $) & (${10}^{3}$AU) & (${10}^{3}$AU) & (\rm Jy/beam) & (Jy) & $({10}^{-18}g/{cm}^{3})$ & $({10}^{5}/{cm}^{3})$ & $ (K) $ &(M$_\odot$) & \\
\hline

Ophcso350-58 &  16h27m22.1s   &  -24d40m59.2s &  28.8 &  14.3 &  10.6 &   2.7 &   4.0 &   0.9 &   5.2 &   1.9 &   5.0 &  21.9 &  0.26 & SP \\
Ophcso350-59 &  16h27m24.2s   &  -24d40m56.7s &  30.8 &  20.6 &  87.8 &   3.3 &   3.2 &   1.9 &  17.2 &   3.1 &   8.0 &  22.6 &  0.78 & PR \\
Ophcso350-60 &  16h27m24.2s   &  -24d27m45.6s &  47.8 &  24.0 & 165.8 &   4.4 &   4.0 &   2.4 &  39.8 &   2.3 &   6.0 &  25.5 &  1.41 & SL \\
Ophcso350-61 &  16h27m25.1s   &  -24d26m52.2s &  27.8 &  17.2 &  31.5 &   2.9 &   4.8 &   1.4 &   9.3 &   1.7 &   4.4 &  27.6 &  0.28 & SP \\
Ophcso350-62 &  16h27m27.7s   &  -24d16m50.3s &  21.2 &  15.6 &  89.6 &   2.4 &   3.8 &   1.1 &   5.4 &   2.3 &   6.1 &  23.6 &  0.22 & SP \\
Ophcso350-63 &  16h27m28.1s   &  -24d27m 7.5s &  29.3 &  16.0 &   1.4 &   2.8 &   2.7 &   3.3 &  22.8 &   5.0 &  13.1 &  25.4 &  0.81 & PR \\
Ophcso350-64 &  16h27m29.3s   &  -24d27m41.8s &  20.9 &  14.9 & 135.1 &   2.3 &   2.1 &   4.5 &  19.9 &   8.1 &  21.2 &  25.4 &  0.71 & PR \\
Ophcso350-65 &  16h27m32.3s   &  -24d27m13.8s &  48.3 &  14.6 &  72.1 &   3.5 &   4.9 &   1.3 &  13.5 &   1.5 &   4.0 &  26.0 &  0.46 & SP \\
Ophcso350-66 &  16h27m32.5s   &  -24d26m27.9s &  37.9 &  21.5 & 168.9 &   3.7 &   3.8 &   1.8 &  21.3 &   2.3 &   6.1 &  23.9 &  0.86 & SP \\
Ophcso350-67 &  16h27m34.7s   &  -24d26m14.9s &  36.7 &  16.1 &  53.3 &   3.2 &   3.1 &   2.0 &  17.6 &   3.3 &   8.8 &  23.1 &  0.77 & SL \\
Ophcso350-68 &  16h27m43.7s   &  -24d42m31.0s &  24.5 &  19.8 & 119.4 &   2.9 &   4.5 &   0.8 &   5.7 &   1.5 &   4.0 &  22.5 &  0.26 & SP \\
Ophcso350-69 &  16h27m52.0s   &  -24d31m41.8s &  26.4 &  17.9 &  85.9 &   2.8 &   3.7 &   1.0 &   7.1 &   2.2 &   5.8 &  21.5 &  0.36 & PR \\
Ophcso350-70 &  16h27m55.1s   &  -24d41m41.1s &  31.4 &  19.5 &  11.0 &   3.2 &   4.7 &   0.8 &   7.6 &   1.4 &   3.8 &  22.6 &  0.35 & SP \\
Ophcso350-71 &  16h27m58.6s   &  -24d33m32.8s &  35.9 &  20.1 &  53.1 &   3.5 &   3.2 &   1.6 &  16.9 &   2.9 &   7.6 &  21.2 &  0.89 & SL \\
Ophcso350-72 &  16h27m60.0s   &  -24d32m57.3s &  32.4 &  28.1 &  31.5 &   4.0 &   3.9 &   1.2 &  16.2 &   2.0 &   5.1 &  21.2 &  0.86 & SL \\
Ophcso350-73 &  16h28m17.0s   &  -24d35m19.3s &  28.0 &  15.5 & 106.4 &   2.7 &   4.0 &   1.1 &   6.9 &   2.0 &   5.4 &  23.2 &  0.30 & SP \\
Ophcso350-74 &  16h28m18.7s   &  -24d35m41.9s &  19.6 &  16.6 &  92.0 &   2.4 &   3.7 &   1.1 &   5.3 &   2.4 &   6.3 &  23.3 &  0.23 & SP \\
Ophcso350-75 &  16h28m19.4s   &  -24d32m58.7s &  29.3 &  21.4 & 161.7 &   3.3 &   4.9 &   0.9 &   8.3 &   1.4 &   3.6 &  23.5 &  0.34 & SP \\

\bottomrule[0.65pt] 
\end{tabular*}
\end{center}

\vspace*{-1mm} {\footnotesize {Notes:}\\
a) {RA and DEC are the positions of the core centroid. The cores are sorted from east to west.}\\
b) {The spatial coverage of each core is ``Ellipse'' in 2D flux density map. ${L}_{maj}$ is the major axis of ellipse. ${L}_{min}$  is the minor axis of ellipse. The values of them are equal to the full width at half maximum (FWHM) of the fitted Gaussian. And they are corrected to remove the effect of the instrumental beam width. }\\
c) {The angle in the clockwise direction from north to the major axis of the ellipse.}\\
d) {The radius of the core. The relationship between mass and radius is shown in Fig. 6. }\\
e) {${\lambda }_{J}$ is the Jeans Length. The distribution of the cores' Jeans Lengths is shown in Fig. 5. }\\
f) {The peak value in the core, which is also corrected to remove the effect of the instrumental beam width.}\\
g) {The core's total flux in the $350 \mu m$ band is derived from GaussClumps.}\\
h) {$\rho$ is the core's mean mass density. n is the core's mean number density. We assume a spherical shape for a core, $V=(4/3)\pi{R}^{3}$. Then we can calculate $\rho$ using the equation: $\rho =m/V $. And $n=(M/\mu)/V$, $\mu$ is the mass per molecular ($\mu =2.3g/{N}_{A}$. ${N}_{A}$ is the Avogadro constant). }\\
i) {The excitation temperature.}\\
j) {The core's total mass.}\\
k) {The type of the cores. SL: starless core. SP: prestellar core. PR: protostellar core. Prestellar cores are a subcategory of starless cores. }

}
\end{table}

\begin{table}[p]
\label{tab:table1}
\caption{Parameters used in GaussClumps \hspace*{3mm}}

\begin{center}\footnotesize \doublerulesep 0.2pt \tabcolsep 5pt
\begin{tabular*}{\textwidth}{ccccccccccccc}\toprule[0.65pt]  
Parameter&FwhmBeam&ExtraCols&FwhmStart&MaxBad&MaxNF&MaxSkip&RMS&Thresh&ModelLim&NPad&Wwidth&Wmin\\
 \hline
Value&5.25&1&3.4&0.05&200&50&0.0364853&10&0.05&50&2&0.05\\
\hline
\bottomrule[0.65pt] 
\end{tabular*}
\end{center}

  \vspace*{-1mm} {\wuhao{Notes:}}\\
{\wuhao FwhmBeam: }{\footnotesize FWHM of the instrument beam, in pixels. Fitted Gaussians are not allowed to be smaller than the instrument beam.}\\
{\wuhao ExtraCols: }{\footnotesize ExtraCols is set to 1, then the catalogue will include columns with names GCFWHM $<$ i $>$ (where $<$ i $>$ is 1, 2), containing the FWHM of the fitted Gaussian in units of pixels. These FWHM values have not been reduced to exclude the effect of the beam width. If the core’s GCFWHMs are less than two beam sizes, the cores are unconvincing and are removed.}\\
{\wuhao FwhmStart: }{\footnotesize An initial guess at the ratio of the typical observed clump size to the instrument beam width. This is used to determine the starting point for the algorithm which finds the best fitting Gaussian for each core.}\\
{\wuhao MaxBad: }{\footnotesize Maximum fraction of bad pixels which may be included in a core. Cores will be excluded if they contain more bad pixels than this value.}\\
{\wuhao MaxNF: }{\footnotesize Maximum number of evaluations of the objective function allowed. We set a large value in order to fit every core well.}\\
{\wuhao MaxSkip: }{\footnotesize  Maximum number of consecutive failures which are allowed when fitting Guassians.}\\
{\wuhao RMS: }{\footnotesize Global RMS (root mean square) noise level of our background-subtracted flux density map.}\\
{\wuhao Thresh: }{\footnotesize Gives the minimum peak amplitude of cores to be fitted by the GaussClumps algorithm. The supplied value is multipled by the RMS noise level before being used. }\\
{\wuhao ModelLim: }{\footnotesize Gaussian model values below the value of the ModelLim multiplied by the RMS noise are treated as zero. }\\
{\wuhao  NPad: }{\footnotesize Algorithm will terminate when ``Npad" consecutive cores have been fitted, all of which have peak values less than the ``Thresh" value.}\\
{\wuhao   Wwidth: }{\footnotesize Ratio of the width of the Gaussian weighting function  to the width of the initial guess Guassian. If this value is too big, a lot of cores will be omitted, and the calculation speed will be very slow.}\\
{\wuhao   Wmin: }{\footnotesize Pixels with weight smaller than this value are not included in the fitting process.}\\
\end{table}

\begin{multicols}{2}

\begin{figure}[H]
\centering
\vspace{10mm}
\hspace*{-7mm}
\includegraphics[scale=0.32,angle=0]{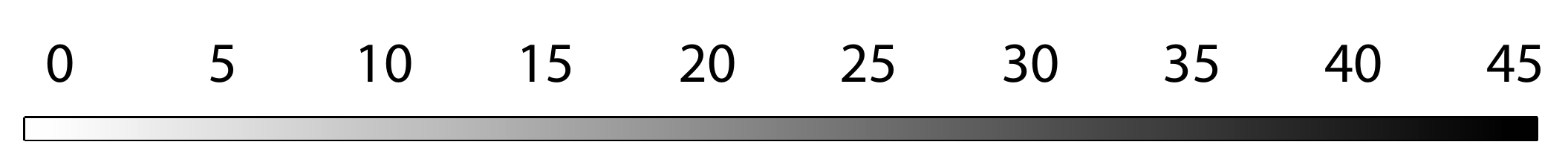}\vspace{3mm}\\
\vspace{-10mm}
\hspace*{-15mm}
\includegraphics[scale=0.38,angle=0]{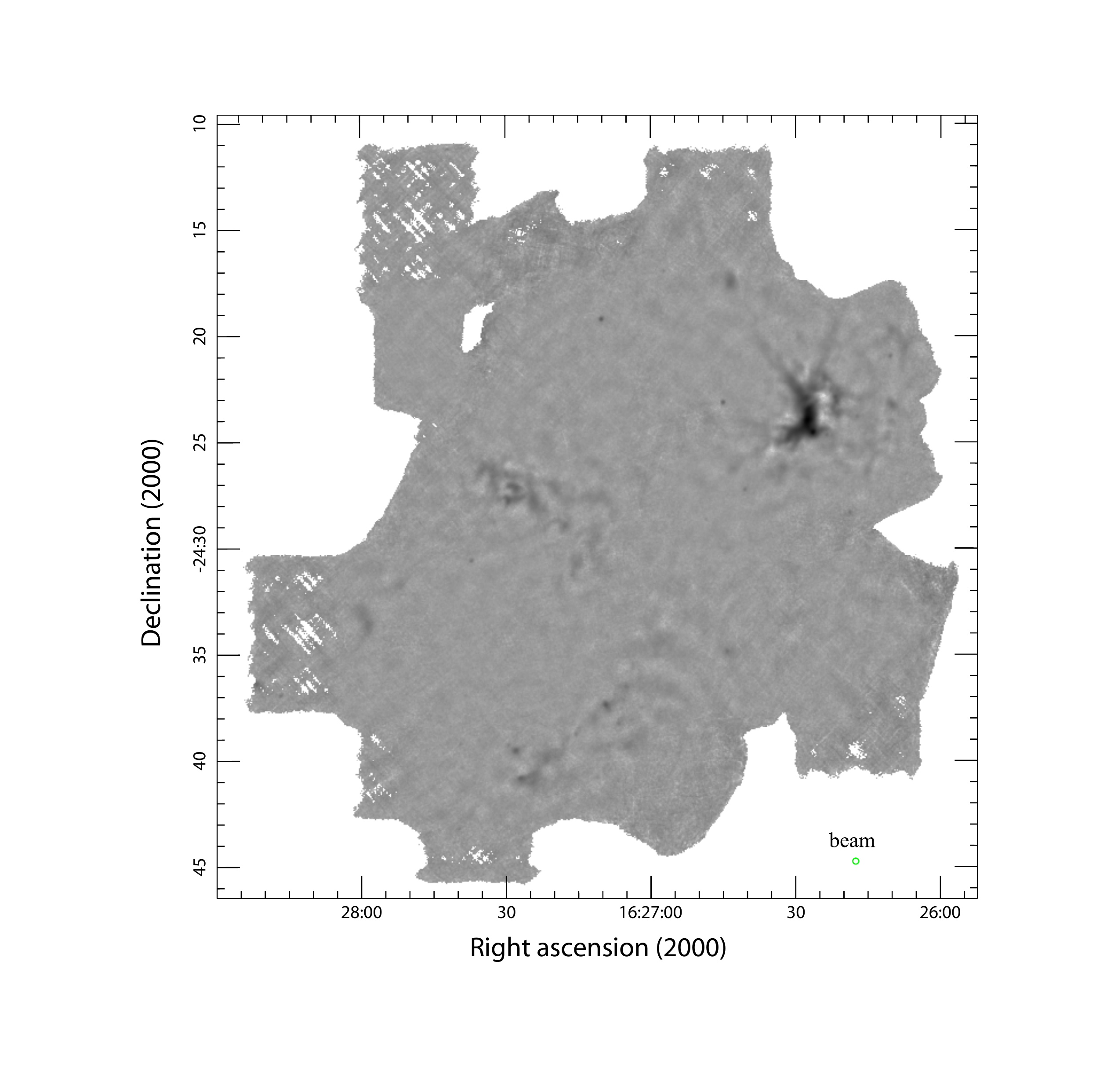}
\vspace{-10mm}
\caption{Ophiuchus molecular cloud flux density map, tracing the dust continuum at 350 $\mu m$ plotted in units of Jy per 8.5 \arcsec\ beam coded as shown in the color bar on the top. Coverage is approximately 0.25 \squaredeg, with 1.61833 arcseconds per pixel. All pixels with an integration time of less than 2 seconds were replaced by NaN (no finite value). Background was subtracted with the FINDBACK procedure.}
\end{figure}

\begin{figure}[H]
\centering
\vspace{10mm}
\hspace*{-11mm}
\includegraphics[scale=0.32,angle=0]{mapbar.pdf}\vspace{3mm}\\
\vspace{-10mm}
\hspace*{-15mm}
\begin{minipage}{0.6\linewidth}
\includegraphics[scale=0.38,angle=0]{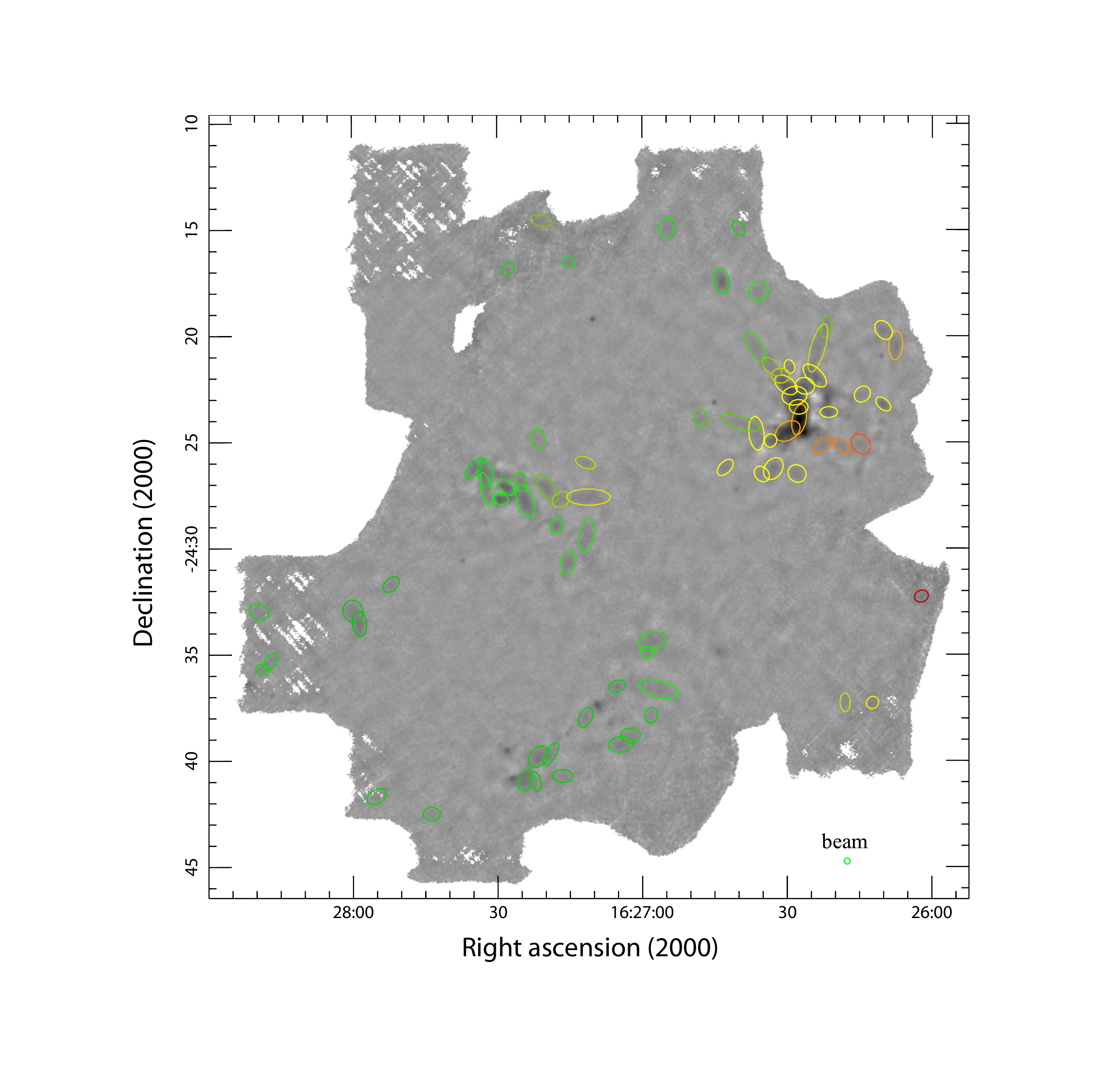}
\end{minipage}
\hfill
\begin{minipage}{0.09\linewidth}
\vspace{-10mm}
\hspace*{-5mm}
\includegraphics[scale=0.32,angle=90]{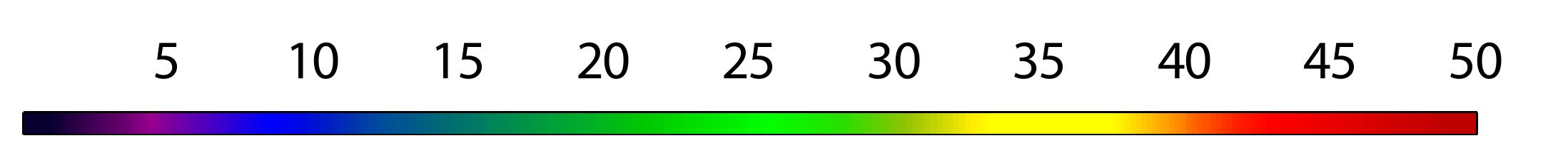}\vspace{0mm}
\end{minipage}
\vspace{-10mm}
\caption{Samplig of 75 cores were detected in the Ophiuchus molecular cloud flux intensity map, and are displayed with measured excitation temperature in units of K, coded as shown in the color bar on the right. Semi-axis of the ellipse plotted in the map is the FWHM of the Gaussian profile of the core, which have been reduced to exclude the effect of the beam width.}
\end{figure}

\end{multicols}
$\ $
\begin{multicols}{2}

\begin{figure}[H]
\centering
\vspace{10mm}
\hspace*{-11mm}
\includegraphics[scale=0.32,angle=0]{mapbar.pdf}\vspace{3mm}\\
\vspace{-10mm}
\hspace*{-15mm}
\begin{minipage}{0.6\linewidth}
\includegraphics[scale=0.38,angle=0]{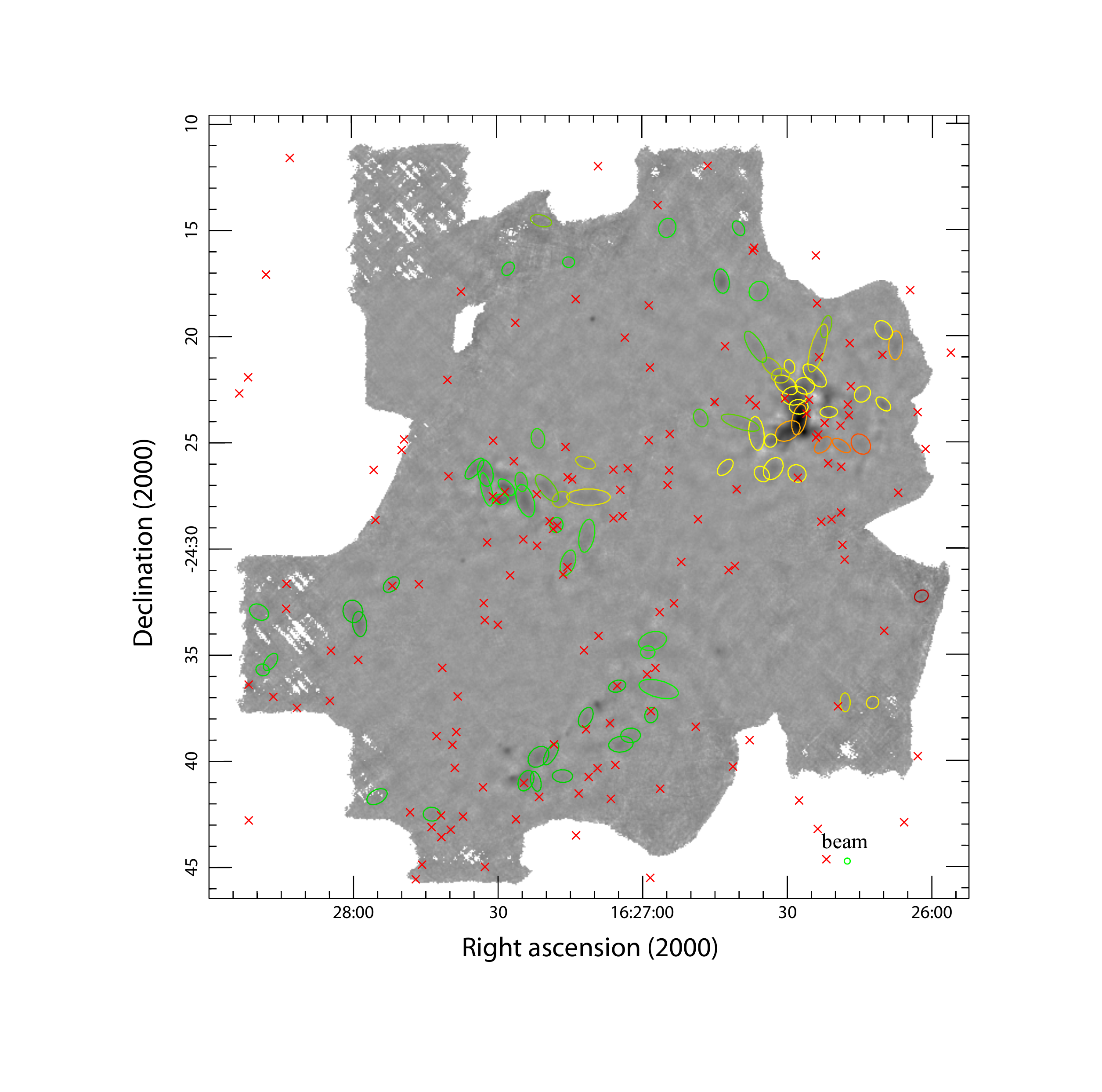}
\end{minipage}
\hfill
\begin{minipage}{0.09\linewidth}
\vspace{-10mm}
\hspace*{-5mm}
\includegraphics[scale=0.32,angle=90]{tempbar.pdf}\vspace{0mm}
\end{minipage}
\vspace{-10mm}
\caption{Sampling of 147 Young Stellar Objects (YSO) are indicated with a red cross symbol. These were selected from the Spitzer c2d candidate YSO clouds catalog of the Ophiuchus molecular cloud. }
\end{figure}

\begin{figure}[H]
\centering
\vspace{10mm}
\hspace*{-11mm}
\includegraphics[scale=0.32,angle=0]{mapbar.pdf}\vspace{3mm}\\
\vspace{-10mm}
\hspace*{-15mm}
\begin{minipage}{0.6\linewidth}
\includegraphics[scale=0.38,angle=0]{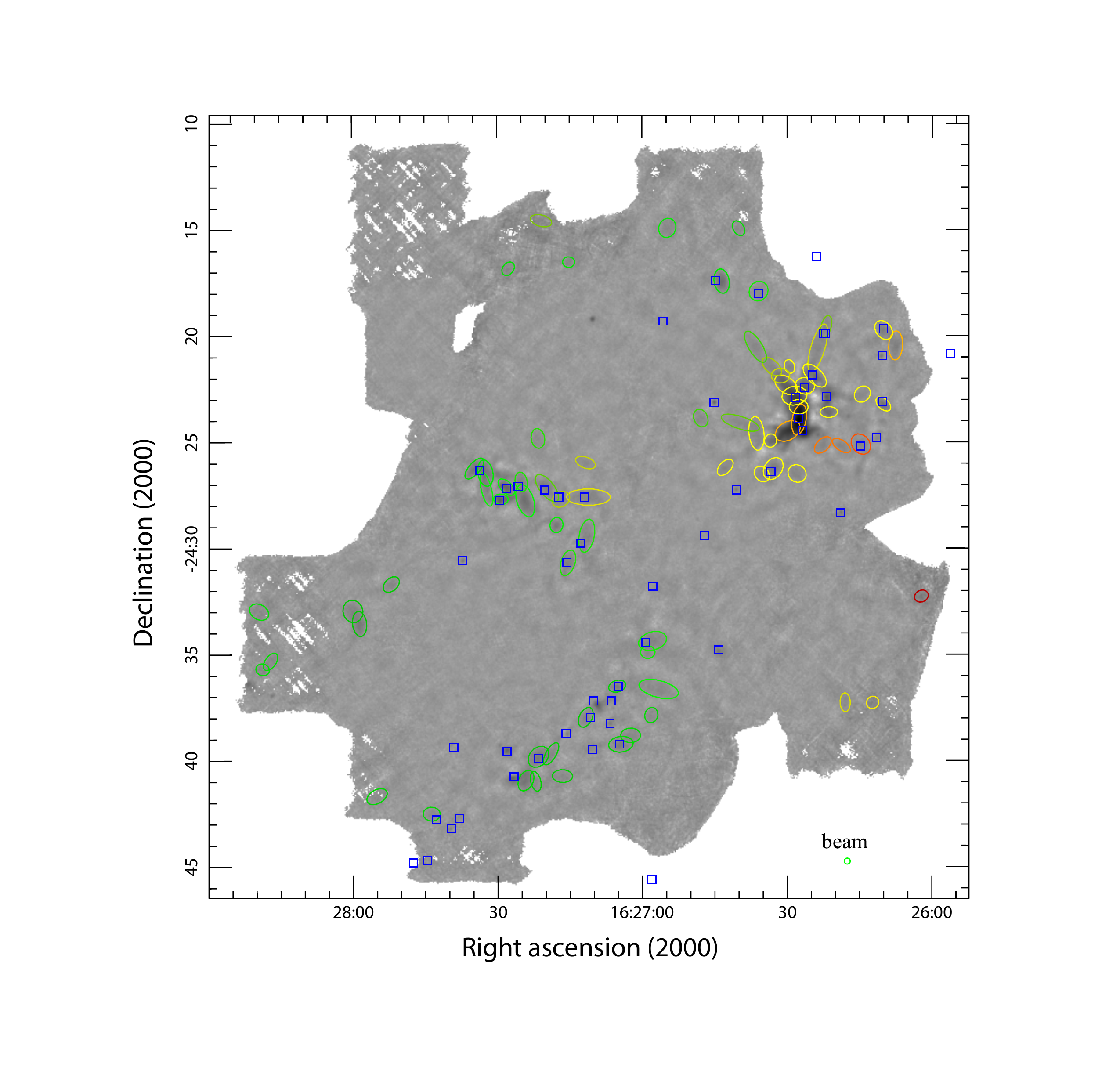}
\end{minipage}
\hfill
\begin{minipage}{0.09\linewidth}
\vspace{-10mm}
\hspace*{-5mm}
\includegraphics[scale=0.32,angle=90]{tempbar.pdf}\vspace{0mm}
\end{minipage}
\vspace{-10mm}
\caption{Sampling of 54 cores detected by Johnstone et al.(2000) at 850 \micron\  are indicated with a blue square symbol. }
\end{figure}

is similar to that used in Li et al. [16], to fit the cumulative functional forms of the CMF. First we generated a random sample of cores according to each of the above distributions. For the power law distribution, we needed to determine three parameters: $\rm {M}_{max},{M}_{min} \ and \ \alpha$ . At low mass end, the sample is incomplete. To reduce the influence of this incompleteness, we set the $\rm {M}_{max}$ value of core mass to the maximum value of the sample. However, for the log-normal distribution, we needed to determine two parameters: $\sigma$ and $\mu $. We first used cumulative functional form of the CMF to generate a random sample: $\rm C({M}_{i}),\ i=1, ...,n$. Then we randomly selected a sample: $\rm {C}_{0}({M}_{i}),\ i=1, ...,n$, from the cores observed in this work. Finally we determined parameters through minimizing the following ${\chi }^{2}$ function:
\begin{equation}
{\chi }^{2}\ \equiv\ \sum_{n}^{i=1}\frac{{[C({M}_{i})-{C}_{0}({M}_{i})]}^{2}}{{C}_{0}{({M}_{i})}^{2}}.
\end{equation}
The results are shown in Fig. 10. We set $\rm {M}_{max}=10.0\, M_\odot$, $\rm {M}_{min}=0.1\, M_\odot$. Based on the power law functional form, we used five different single-index $(\alpha)$: -1.9, -1.5, -1.0, -0.8, and -0.4 to simulate core samples. As the $(\alpha)$ value increases, the shape of the CMF tends towards a straight line. Simulated core samples are displayed in Fig. 9.
\begin{figure}[H]
\begin{center}
\includegraphics[scale=0.5,angle=0]{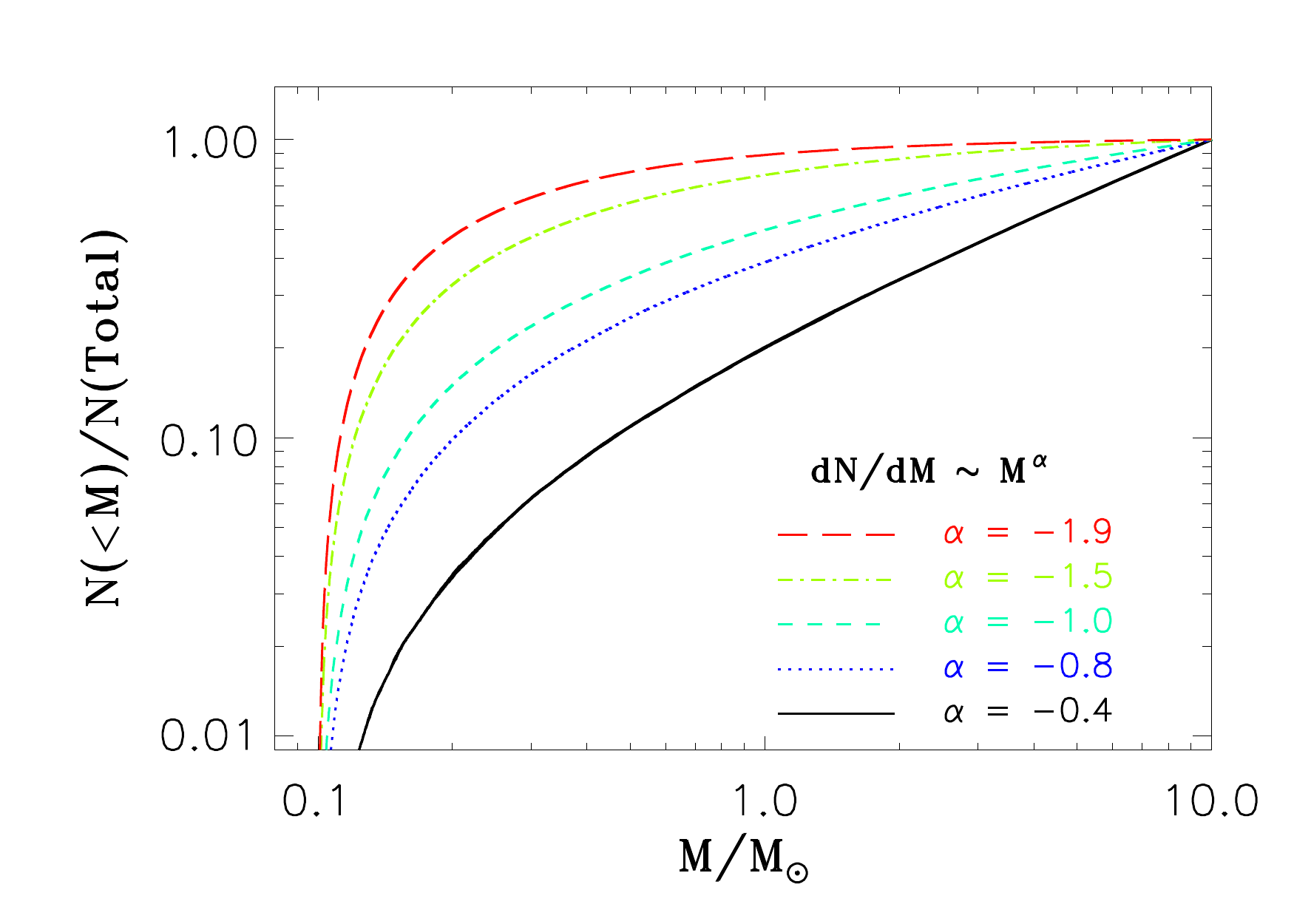}
\caption{Cumulative CMFs of simulated core samples, based on the five different single-index fits, are plotted with different line styles in five colors.}
\end{center}
\end{figure}
\end{multicols}

\begin{figure}[H]
\begin{minipage}{0.5\linewidth}
  \centerline{\includegraphics[scale=0.4]{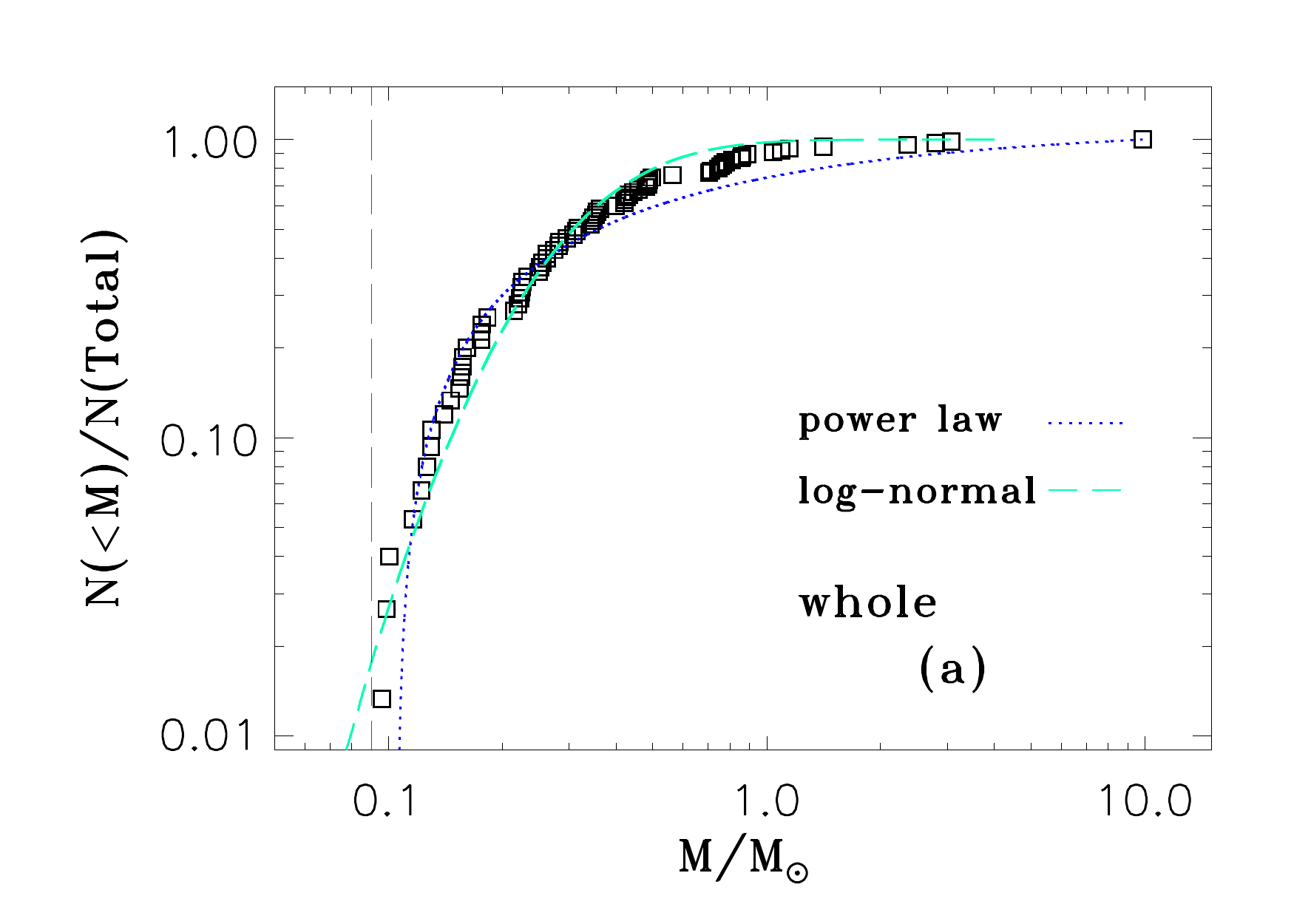}}
\end{minipage}
\hfill
\begin{minipage}{0.5\linewidth}
  \centerline{\includegraphics[scale=0.4]{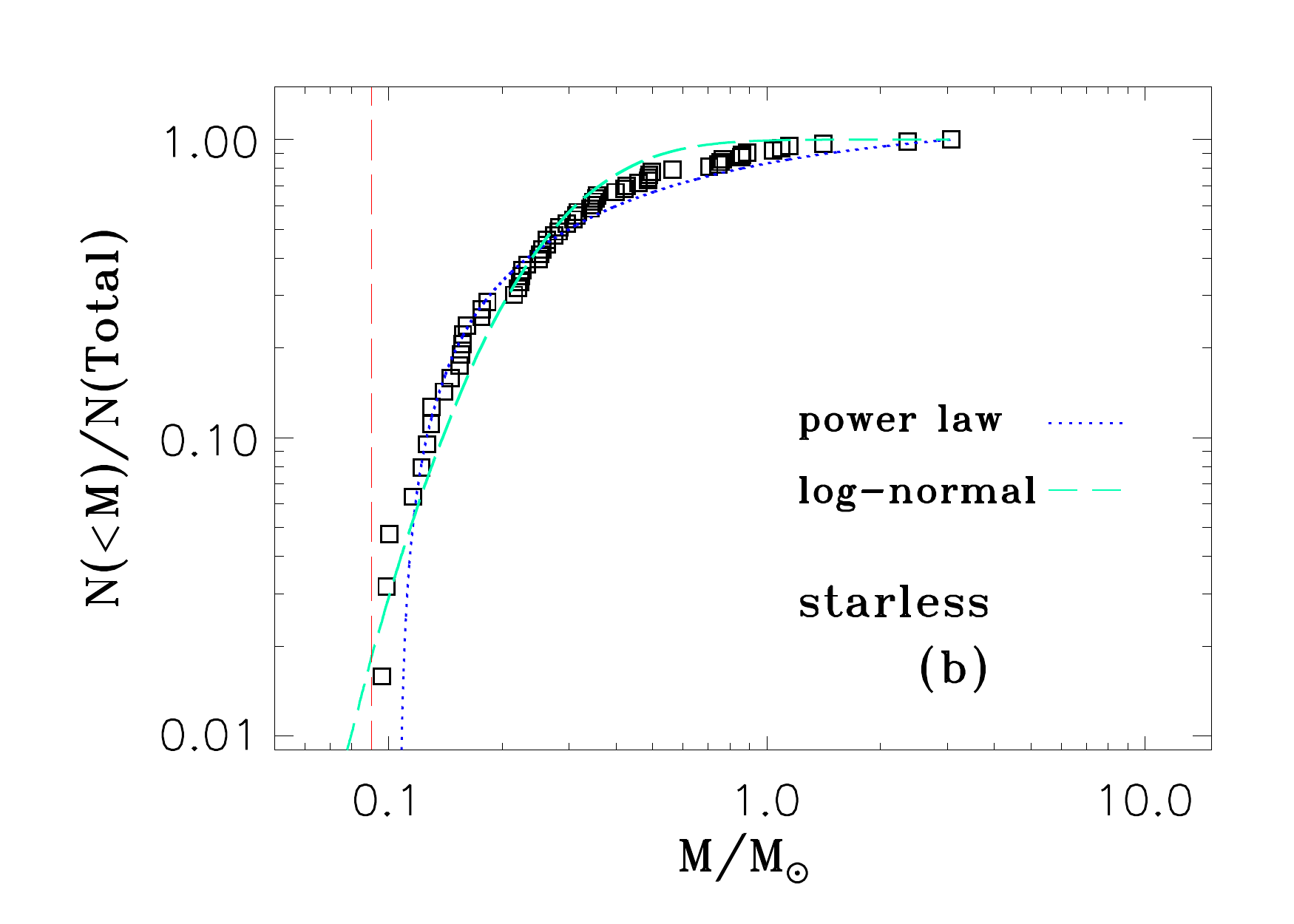}}
\end{minipage}
\vfill
\begin{minipage}{0.5\linewidth}
  \centerline{\includegraphics[scale=0.4]{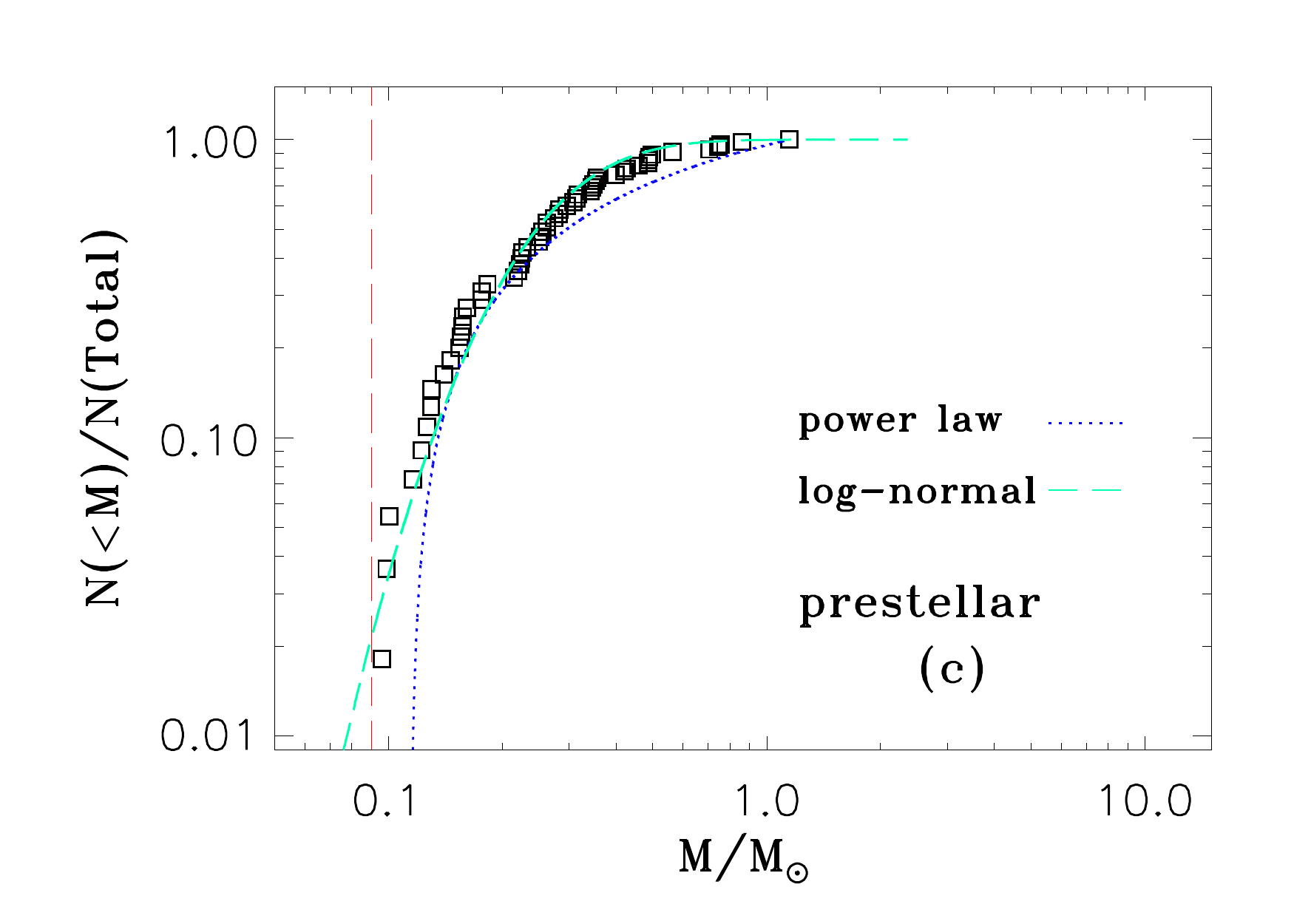}}
\end{minipage}
\hfill
\begin{minipage}{0.5\linewidth}
  \centerline{\includegraphics[scale=0.4]{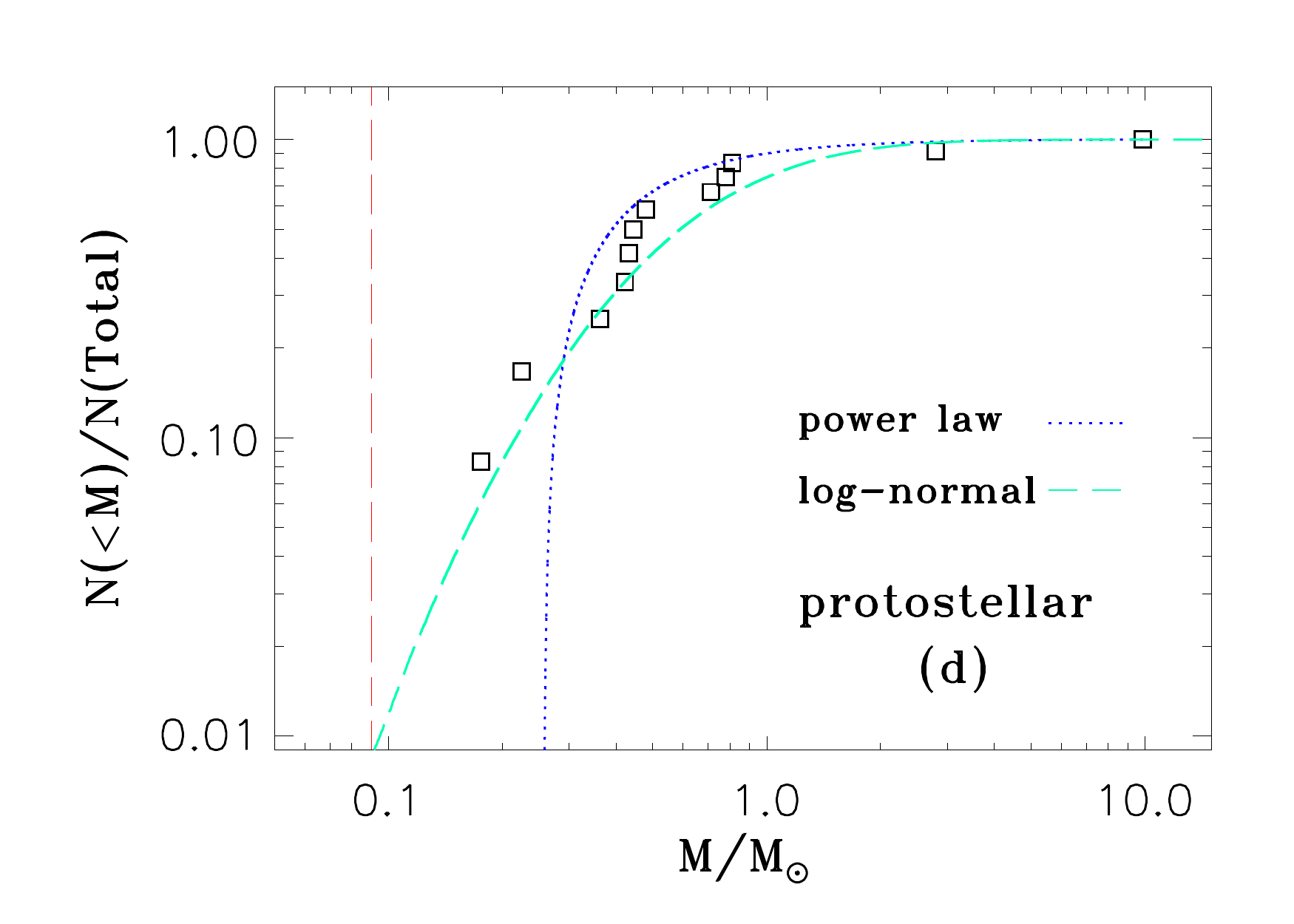}}
\end{minipage}

\caption{Mass distribution of the cores found by GaussClumps fitting to the Ophiuchus molecular cloud flux density map at $350\ \mu m$ . Cores found by GaussClumps fitting are indicated by the square symbols. Power law CMFs are plotted in a blue dotted line, while the log-normal CMFs are plotted in a cyan long dashes line. Red lines on the figure represent the upper mass boundary of the region of incompleteness, 0.09 ${M}_{\odot}$ as discussed in section 3.4. In plot (a) we use whole cores to fit the CMF. Fitted results are as follows: A power law with $dN/dM\sim {M}^{-1.48\pm0.13 },\ {M}_{max}=9.9M_\odot,\ {M}_{min}=0.11\pm0.01M_\odot$, and a log-normal distribution with $\mu =-1.18\pm0.10,\ \sigma =0.58\pm0.05$.
In plot (b) we use starless cores to fit the CMF. Fitted results are as follows: A power law with $dN/dM\sim {M}^{-1.51\pm0.20},\ {M}_{max}=3.1M_\odot,\ {M}_{min}=0.11\pm0.01M_\odot$, and a log-normal distribution with $\mu =-1.30\pm0.11,\ \sigma =0.53\pm0.05$.
In plot (c) we use prestellar cores to fit the CMF. Fitted results are as follows: A power law with $dN/dM\sim {M}^{-1.31\pm0.28 },\ {M}_{max}=1.1M_\odot,\ {M}_{min}=0.11\pm0.09M_\odot$, and a log-normal distribution with $\mu =1.40\pm0.10,\ \sigma =0.50\pm0.05$.
In plot (d) we use protostellar cores to fit the CMF. Fitted results are as follows: $dN/dM\sim {M}^{-2.66\pm2.63 },\ {M}_{max}=9.9M_\odot,\ {M}_{min}=0.26\pm0.13M_\odot$, and a log-normal distribution with $\mu =-0.53\pm0.30,\ \sigma =0.79\pm0.26$.}

\end{figure}

\begin{multicols}{2}

\subsection{Power law}
The starless CMF can be well fitted by the power law $\ $form. But the exponent ($-1.51\pm 0.20$) of the starless CMF is greater than the exponent of the Salpeter IMF ($\alpha =-2.35$). Clark et al. [45] suggested that the CMF depends on the core lifetime, which may be why the exponent of the starless CMF is greater than the exponent of the Salpeter IMF. The starless CMF is also related to the process of star formation and the IMF. The process of star formation determines the speed of the evolution of the starless core at a certain mass. The IMF determines the relative number of stars after the star formation is complete [46]. At the low mass end, sensitivity limits cause a reduction in the number of cores found. This completeness affects the shape of the CMF.
\subsection{Log-normal}
Turbulence within molecular clouds is highly supersonic, which plays a major role in the process of star formation [47-49]. In molecular clouds, large-scale supersonic turbulence cascades to small scales through shocks [50]. The probability density function (PDF) [51] can describe the gas density fluctuations created by turbulence. A log-normal PDF was predicted to arise when the central limit theorem is applied to isothermal turbulence [13,14]. This has been reproduced in computer simulations [52,53]. A log-normal PDF was naturally predicted, when the central limit theorem is applied to isothermal turbulence [13,14]. And this has been produced in computer simulations [52,53]. Padoan and Nordlund [12] first derived the mass distribution of gravitationally unstable cores with a log-normal PDF of the mass density in supersonic turbulence, and concluded that turbulent fragmentation is essential to the origin of the stellar IMF. Such a log-normal behaviour for density fluctuations has received observational support. In this work, the whole and prestellar CMF are both well fitted by log-normal distributions. This finding $\ $suggests that turbulence influences the evolution of the Ophiuchus molecular cloud.\\

\section{Conclusions}
Using the SHARC-II camera at the CSO telescope, we obtained dust continuum data at 350 \micron\ for the Ophiuchus molecular cloud. These data were mainly used to determine the CMF of the Ophiuchus molecular cloud. The total masses, mean temperatures, Jeans Length, and mean mass density of the cores were also calculated.\\
We summarize the primary results of the paper as follows:\\
Firstly, A 350 \micron\ map covering 0.25 deg${}^{2}$ of the Ophiuchus molecular cloud was created by mosaicing 56 separate scans.\\
Secondly, This map of the Ophiuchus molecular cloud was analyzed using the GaussClumps algorithm, in which 75 cores have been detected. The mean temperature of the cores was 29K and the mean mass of the cores was 0.6 M$_\odot$. \\
Thirdly, We used the Spitzer c2d catalogs to separate 63 starless cores from 12 protostellar cores. 55 prestellar cores were also distinguished by their property of Jeans instability.\\
Fourthly, We performed a comparison between the core catalogue at 350 \micron\ and the JCMT/SCUBA 850 \micron\ catalogue. The cores unique to the 350 \micron\ map may have too high a temperature to be detected by the 850 \micron\ survey. Some of the cores apparently missing from this 350 \micron\ study are likely to be cooler
sources.\\
Lastly, We found that the whole and prestellar CMF can be well fitted by log-normal distributions with $\mu =-1.18\pm0.10,\ \sigma =0.58\pm0.05$ and $\mu =1.40\pm0.10,\ \sigma =0.50\pm0.05$  respectively, while the starless CMF can be better fitted by a power law: $\rm dN/dM \sim {M}^{-1.51\pm 0.20},\ {M}_{max}=3.1\ M_\odot,\ {M}_{min}=0.11\pm 0.01\ M_\odot$.
\vspace*{2mm}
\Acknowledgements{\bahao This material is based upon work at the Caltech Submillimeter Observatory, which is operated by the California Institute of Technology under cooperative agreement with the National Science Foundation (AST-0838261). This work was supported by National Basic Research Program of China (Grant Nos. 2012CB821800), National Aeronautics and Space Administration Undergraduate Student Research Program of USA, Natural Science Foundation of China (Grant Nos. 11373038 and 11163002), Graduate Innovative Fund of GuiZhou University (Grant Nos. 2013024).\\
This research was carried out at the NAOC ISM Group. We would to thank M. Tafalla for useful discussions on this paper. We thank D.S. Berry and P.W. Draper for their technical support with STARLINK. Special acknowledgement is also given to A.Y. Zhou. }
\normalsize \vskip0.1in\parskip=0mm \baselineskip 18pt
\renewcommand{\baselinestretch}{1.06}\footnotesize\parindent=4mm\bahao

\vskip0.1in \noindent 
\vskip0.1in\parskip=0mm

\REF{1\ }Ward-Thompson D, Scott P F, et al. A submillimetre continuum survey of pre protostellar cores. Mon Not R Astron Soc, 1994, 268: 276

\REF{2\ }Kirk J M, Ward-Thompson D, Andr$\acute{e}$ P. The initial conditions of isolated star formation - VI. SCUBA mapping of pre-stellar cores. Mon Not R Astron Soc, 2005, 360: 1506-1526

\REF{3\ }Ward-Thompson D, Andre P, Crutcher R, et al. Protostars and Planets V, Edited by Reipurth B, Jewitt D, \& Keil K. The University of Arizona Press, 2007

\REF{4\ }Alves J, Lombardi M, Lada C J. The mass function of dense molecular cores and the origin of the IMF. Astron Astrophys, 2007, 462:L17--L21

\REF{5\ }Lada C J, Muench A A, et al. The Nature of the Dense Core Population in the Pipe Nebula: Thermal Cores Under Pressure. Astrophys J, 2008, 672: 410--422

\REF{6\ }Evans II N J, Dunham M M, J$\phi $rgensen J K, et al. The Spitzer c2d Legacy Results: Star-Formation Rates and Efficiencies; Evolution and Lifetimes. Astrophys J Suppl, 2009, 181: 321--350

\REF{7\ }Tafalla M. Molecules in outflows from Young Stellar Objects. Astronomical Society of the Pacific. 2013, 476: 177

\REF{8\ }Tafalla M, Hacar A. HH 114 MMS: a new chemically active outflow. Astron Astrophys. 2013, 552: L9

\REF{9\ }Tafalla M, Liseau R, et al. High-pressure, low-abundance water in bipolar outflows. Results from a Herschel-WISH survey. Astron Astrophys. 2013, 551: A116

\REF{10\ }Enoch M, Evans II N, Sargent A, et al. The Mass Distribution and Lifetime of Prestellar Cores in Perseus, Serpens, and Ophiuchus. Astrophys J, 2008, 684: 1240--1259

\REF{11\ }Ballesteros-Paredes J, Gazol A, Kim J, et al. The Mass Spectra of Cores in Turbulent Molecular Clouds and Implications for the Initial Mass Function. Astrophys J, 2006, 637:384--391

\REF{12\ }Padoan P and Nordlund $\rm \r A$. The Stellar Initial Mass Function from Turbulent Fragmentation. Astrophys J, 2002, 576: 870--879

\REF{13\ }Larson R B. Dynamical Models for the Formation and Evolution of Spherical Galaxies. Mon Not R Astron Soc, 1973, 5: 320

\REF{14\ }Adams F C, Fatuzzo M. A Theory of the Initial Mass Function for Star Formation in Molecular Clouds. Astrophys J, 1996, 464: 256

\REF{15\ }Swift J J, Beaumont C N. Discerning the Form of the Dense Core Mass Function. Publ Astron Soc Pac, 2010, 122: 224--230

\REF{16\ }Li D, Velusamy T, Goldsmith P, et al. Massive Quiescent Cores in Orion. II. Core Mass Function. Astrophys J, 2007, 655: 351--363

\REF{17\ }Casassus S, Dickinson C, et al. Centimetre-wave continuum radiation from the ρ Ophiuchi molecular cloud. Mon Not R Astron Soc, 2008, 391: 1075--1090

\REF{18\ }Motte F, Andre, P, \& Neri, R. The initial conditions of star formation in the $\rho$  Ophiuchi main cloud: wide-field millimeter continuum mapping. Astron Astrophys, 1998, 336: 150--172

\REF{19\ }Johnstone D, Wilson C, Moriarty-Schieven G, et al. Large-Area Mapping at 850 Microns. II. Analysis of the Clump Distribution in the $\rho$ Ophiuchi Molecular Cloud. Astrophys J, 2000, 545: 327--339

\REF{20\ }Young K E, et al. Bolocam Survey for 1.1 mm Dust Continuum Emission in the c2d Legacy Clouds. II. Ophiuchus. Astrophys J, 2006, 644: 326--343

\REF{21\ }Stanke T, Smith M, Gredel R, et al. An unbiased search for the signatures of protostars in the ρ Ophiuchi molecular cloud . II. Millimetre continuum observations. Astron Astrophys, 2006, 447: 609--622

\REF{22\ }Pilbratt G L. et al. Herschel Space Observatory. An ESA facility for far-infrared and submillimetre astronomy. Astron Astrophys, 2010, 518

\REF{23\ }Dowell C D, et al. SHARC II: a Caltech Submillimeter Observatory facility camera with 384 pixels. Proc SPIE, 2003, 4855: 73--87

\REF{24\ }$\rm Kov{\acute{a}}cs$ A. Scanning strategies for imaging arrays. Proc Int Soc Opt Photon, 2008, 7020

\REF{25\ }$\rm Kov{\acute{a}}cs$ A. CRUSH: fast and scalable data reduction for imaging arrays. Int Soc Opt Photon, 2008, 7020

\REF{26\ }Berry D, Currie M, et al. Starlink 2012: The Kapuahi Release, Astronomical Society of the Pacific, 2013, 247

\REF{27\ }Stutzki J and Guesten R. High spatial resolution isotopic CO and CS observations of M17 SW - The clumpy structure of the molecular cloud core. Astrophys J, 1990, 356: 513--533

\REF{28\ }Lombardi M, Bertin G. Boyle's law and gravitational instability. Astron Astrophys. 2001, 375: 1091--1099

\REF{29\ }Goldsmith P F. Molecular depletion and thermal balance in dark cloud cores. Astrophys J. 2001, 557: 736--746

\REF{30\ }Qian L, Li D, Goldsmith P. ${}^{13}$CO Cores in the Taurus Molecular Cloud. Astrophys J, 2012, 760

\REF{31\ }Bontemps S, Andr$\acute{e}$ P, et al. ISOCAM observations of the rho Ophiuchi cloud: Luminosity and mass functions of the pre-main sequence embedded cluster. Astron Astrophys, 2001, 372: 173--194

\REF{32\ }Mamajek E E. On the distance to the Ophiuchus star-forming region. Astron Nachr, 2008, 329: 10

\REF{33\ }Ossenkopf V, \& Henning T. Dust opacities for protostellar cores. Astron Astrophys, 1994, 291:943--959

\REF{34\ }J$\phi $rgensen J K, Johnstone D, et al. Current Star Formation in the Perseus Molecular Cloud: Constraints from Unbiased Submillimeter and Mid-Infrared Surveys. Astrophys J, 2007, 656:293--305

\REF{35\ }Enoch M L, Evans II N J, et al. Properties of the Youngest Protostars in Perseus, Serpens, and Ophiuchus. Astrophys J, 2009, 692:973--997

\REF{36\ }Evans et al. Final Delivery of Data from the c2d Legacy Project: IRAC and MIPS. Pasadena, CA: SSC. 2007

\REF{37\ }Lomax O D. Simulations of star formation in Ophiuchus. PhD thesis .Cardiff university. 2013

\REF{38\ }Sadavoy S I, Di Francesco J, Johnstone D. "Starless" Super-Jeans Cores in Four Gould Belt Clouds. Astrophys J, 2010, 718:L32--L37

\REF{39\ }Scalo J. Fifty years of IMF variation: the intermediate-mass stars. Edited by Corbelli E and Palle F. Astrophys Space Sci Library. 2005, 327:23

\REF{40\ }Salpeter E E. The Luminosity Function and Stellar Evolution. Astrophys J, 1955, 121:161

\REF{41\ }Miller G, Scalo J. The initial mass function and stellar birthrate in the solar neighborhood. Astrophys J Suppl Ser, 1979, 41: 513--547

\REF{42\ }Kroupa P. On the variation of the initial mass function. Mon Not R Astron Soc, 2001, 322: 231--246

\REF{43\ }Chabrier G. Galactic Stellar and Substellar Initial Mass Function. Publ Astron Soc Pac, 2003, 115: 763--795

\REF{44\ }Olmi L, Angl$\acute{e}$s-Alc$\acute{a}$zar D, et al. On the shape of the mass-function of dense clumps in the Hi-GAL fields . I. Spectral energy distribution determination and global properties of the mass-functions. Astron Astrophys, 2013, 551: A111

\REF{45\ }Clark P C, Klessen R S, \& Bonnell I A. Clump lifetimes and the initial mass function. Mon Not R Astron Soc, 2007, 379:57--62

\REF{46\ }McKee C F and Offner S S R. The protostellar mass function. Astrophys J, 2010, 716: 167--180

\REF{47\ }Ballesteros-Paredes J, Klessen R S, et al. Molecular Cloud Turbulence and Star Formation. Protostars and Planets V, Edited by Reipurth B, Jewitt D, \& Keil K. The University of Arizona Press, 2007

\REF{48\ }McKee C and Ostriker, E. Theory of Star Formation. Annual Review of Astronomy and Astrophysics. 2012, 45:565--687

\REF{49\ }Joos M, Hennebelle P, et al. The influence of turbulence during magnetized core collapse and its consequences on low-mass star formation. 2013, 554:A17

\REF{50\ }Chabrier G and  Hennebelle P. Dimensional argument for the impact of turbulent support on the stellar initial mass function. Astron Astrophys, 2011, 534: A106

\REF{51\ }Federrath C, Klessen R S. The star formation rate of turbulent magnetized clouds: comparing theory, simulations, and observations. Astrophys J, 2012,761:156

\REF{52\ }V$\acute{a}$zquez-Semadeni, E. Hierarchical Structure in Nearly Pressureless Flows as a Consequence of Self-similar Statistics. Astrophys J. 1994 ,423:681

\REF{53\ }Klessen R S. The Formation of Stellar Clusters: Mass Spectra from Turbulent Molecular Cloud Fragmentation. Astrophys J, 2001, 556:837--846

\end{multicols}

\end{document}